\UseRawInputEncoding
\pdfoutput=1%For ARXIV
\documentclass{article}

\usepackage{pifont}
\usepackage{graphicx}
\usepackage{multirow} 
\usepackage{caption}
\usepackage{subfigure}

\usepackage[final]{neurips_data_2022}

\usepackage[utf8]{inputenc} % allow utf-8 input
\usepackage[T1]{fontenc}    % use 8-bit T1 fonts
\usepackage{hyperref}       % hyperlinks
\usepackage{url}            % simple URL typesetting
\usepackage{booktabs}       % professional-quality tables
\usepackage{amsfonts}       % blackboard math symbols
\usepackage{nicefrac}       % compact symbols for 1/2, etc.
\usepackage{microtype}      % microtypography
\usepackage{xcolor}         % colors

\usepackage{color}

\title{Tenrec: A Large-scale Multipurpose Benchmark Dataset for Recommender Systems}

\author{
Guanghu Yuan\textsuperscript{\rm 1,3,4}$^*$,~ Fajie Yuan\textsuperscript{\rm 2}$^*$,~Yudong Li\textsuperscript{\rm 1},~Beibei Kong\textsuperscript{\rm 1},~Shujie Li\textsuperscript{\rm 4},~Lei Chen\textsuperscript{\rm 3},\\
\textbf{~Min Yang\textsuperscript{\rm 3},~Chenyun Yu\textsuperscript{\rm 5},~Bo Hu\textsuperscript{\rm 1},~Zang Li\textsuperscript{\rm 1},~Yu Xu\textsuperscript{\rm 1},~Xiaohu Qie\textsuperscript{\rm 1}}\\
\textsuperscript{\rm 1}Tencent\quad
\textsuperscript{\rm 2}Westlake University \\
\textsuperscript{\rm 3}Shenzhen Institute of Advanced Technology, Chinese Academy of Sciences \\
\textsuperscript{\rm 4}University of Science and Technology of China\quad
\textsuperscript{\rm 5}Sun Yat-sen University \\
\texttt{gh.yuan0@gmail.com},
\texttt{yuanfajie@westlake.edu.cn},
\texttt{ustclsj@mail.ustc.edu.cn}\\
\texttt{\{lei.chen,min.yang\}@siat.ac.cn},
\texttt{yuchy35@mail.sysu.edu.cn} \\
\texttt{\{elsonli,echokong,harryyfhu,gavinzli, henrysxu,tigerqie\}@tencent.com}
}

\begin{document}

\maketitle

\def\thefootnote{*}\footnotetext{Equal contribution.  Fajie designed the research, Guanghu performed the research; Fajie, Guanghu, Lei and Min wrote the paper; Fajie, Min, Yudong and Beibei launched  the research project; Guanghu, Beibei and Yudong collected the data; Shujie assisted in performing partial experiments. Experiments of this work were mainly performed when Guanghu interned at Tencent.
}

% \nolinenumbers
\begin{abstract}
Existing benchmark datasets for recommender systems (RS)  either are created  at a small scale or involve very limited forms of user feedback.
% only domain-specific user feedback.
RS models evaluated on such datasets often lack practical values for large-scale real-world applications. In this paper, we describe Tenrec\footnote{Tenrec (a hedgehog-like mammal) here means that the dataset is collected from the recommendation platforms of \underline {Ten}cent, and that it can be used to benchmark \underline {ten} diversified 
recommendation tasks. }, a novel and publicly available data collection for RS that records various user feedback from four different recommendation scenarios. To be specific, Tenrec has the following five characteristics: (1) it is large-scale, containing around 5 million users and 140 million interactions; (2) it has not only positive user feedback, but also true  negative feedback (vs. one-class recommendation); (3) it contains overlapped users and items across four different scenarios; (4) it contains various types of  user positive feedback, in forms of clicks, likes, shares, and follows, etc; (5) it contains additional features beyond the user IDs and item IDs. We verify Tenrec on ten diverse  recommendation  tasks by running several classical baseline models per task. Tenrec has the potential to become a 
useful benchmark dataset for a majority of popular recommendation tasks. 
Our source codes, datasets and leaderboards are available at \textcolor{red}{ \url{https://github.com/yuangh-x/2022-NIPS-Tenrec}}\footnote{ \textcolor{red}{Email Fajie \& Guanghu if you want to launch a new leaderboard for an important RS task using Tenrec.}}. 

\end{abstract}

\section{Introduction}
Recommender systems (RS) aim to estimate user preferences on items that users have not yet seen.  Progress in deep learning (DL) has spawned a wide range of novel and complex neural recommendation models. Many improvements have been achieved in prior literature, however, much of them performs evaluation on non-benchmark datasets or on datasets at a small scale by modern standard. This has led to severe reproducibility and credibility problems in the RS community. For example, ~\cite{rendle2019difficulty} showed that many  `advanced' baselines reported in previous papers are largely suboptimal, even underperforms the vanilla matrix factorization (MF)~\cite{koren2009matrix}, an old baseline proposed over a decade ago. ~\cite{rendle2020neural} further demonstrated that with a 
careful setup dot product is superior to the learned similarity, e.g. using a multilayer perceptron (MLP)~\cite{he2017neural}. Recently, ~\cite{dacrema2019we,krichene2020sampled,sun2020we,ji2020critical,dacrema2021troubling,cremonesi2021progress,anelli2022top} also questioned some recognized progress in RS from the dataset and experimental setup perspectives. 
% , a.k.a. neural collaborative filtering~\cite{he2017neural}.

% With the continuous development of machine learning and deep learning, recommendation systems have made great progress in various scenarios, such as e-commerce websites, news software, and social platforms. And in recent years, recommendation models have also developed from simple and traditional models (ItemKNN[], PMF[]) to complex and advanced models (AFM[], DeepFM[]), and this development has promoted the rapid progress of the recommendation system industry. However, some work in recent years has also questioned the development of current recommender systems. Part of the literature finds that although many works have made progress on some special tasks or special datasets, they are not convincing. Second, some literature finds that there is an inconsistency between the results of sampling items for evaluation and evaluating all items. 
A large-scale and high-quality datasets have a significant impact on accelerating research in an area,
% can significantly benefit a research field,
such as ImageNet~\cite{deng2009imagenet} for computer vision (CV) and GLUE~\cite{wang2018glue} for natural language processing (NLP). 
% and SQuAD~\cite{rajpurkar2016squad} for machine reading comprehension. 
However, it is often difficult for researchers to access large-scale real-world datasets for studying recommendation problems due to security or privacy issues. Despite that, there still exist several popular datasets for regular recommendation tasks. For example, movieLens\footnote{https://grouplens.org/datasets/movielens/} (ML) datasets, including ML-100K, ML-1M and ML-10M, etc have become the stable benchmark datasets for the rating prediction~\cite{koren2009matrix} task.  Other popular datasets, including Netflix~\cite{narayanan2008robust} for movie recommendation,  Yelp~\cite{choi2021empirical} for location recommendation, Amazon~\cite{dwivedi2021product} for product  recommendation,
Mind~\cite{wu2020mind} for news recommendation, Last.fm~\cite{yuan2019simple} and Yahoo! Music~\cite{dror2012yahoo} for song recommendation, also appear frequently  in literature.  One major drawback of these datasets is that user feedback data has very limited forms, for example, most of them  include only one type of user feedback (either rating, or clicking, or watching), or are collected from only one recommendation scenario.  
This severely limits the research scope of real-world recommender systems.
To foster diverse recommendation research, we propose Tenrec, a large-scale and multipurpose real-world  dataset. Compared with existing public datasets, Tenrec has several merits: (1) it consists of overlapped users/items from four different real-world recommendation scenarios, which can be used to study the cross-domain recommendation (CDR) and transfer learning (TF) methods; (2)
it contains multiple types of positive user feedback (e.g. clicks, likes, shares, follows, reads and favorites), which can be leveraged to study the multi-task learning (MTL) problem; (3) it has both positive user feedback and true negative feedback, which can be used to study more practical  click-through rate (CTR)  prediction scenario; (4) it has additional user and item features beyond the identity information (i.e. user IDs and item IDs), which can be used for context/content-based recommendations. 

Owing to these advantages, Tenrec can be employed to evaluate a wide range of recommendation tasks. In this paper, we examine its properties by ten recommendation tasks, including (1) CTR prediction~\cite{zhu2021open, zhu2022bars}, (2) session-based recommendation~\cite{hidasi2015session}, (3) MTL recommendation~\cite{ma2018modeling},  (4) CDR recommendation~\cite{zhu2021cross}, (5) user profile prediction~\cite{yuan2020parameter}, (6) cold-start recommendation~\cite{volkovs2017dropoutnet}, (7) lifelong user representation learning~\cite{yuan2020one}, (8)
model  compression~\cite{sun2020generic}, (9) model training speedup~\cite{wang2020stackrec}, and (10) model inference speedup~\cite{chen2021user}. Beyond these tasks, we can easily integrate some of the above characteristics  to propose additional or new tasks.
To the best of our knowledge, Tenrec is so far one of the largest datasets for RS, 
covering  a majority of recommendation scenarios and  tasks. 
We release all datasets and codes to promote  reproducibility    and advance new recommendation research.

% In this paper, we propose a large-scale multi-scene dataset for recommendation research, to address the above problems, which is collected from Tencent. Unlike previous datasets, it contains more than 6,200,000 users, 3,800,000 items and hundreds of millions of interactions. In particular, it also includes multiple scenarios that can be used for transfer learning, cold start and lifelong learning, etc. 

% Compared with the traditional recommender system dataset, our dataset not only records the user's click behavior, but also records the user's other interaction behavior, which can restore the user's preferences more. At the same time, some characteristics of the user and the item itself are recorded, which can be used to describe the user's profile and item click prediction. 

% We implement many state-of-the-art recommendation methods in different recommender system scenarios, which were originally developed on different proprietary datasets, and compare their method’s performance on our dataset. Experiments show that multi-scenario datasets can help drive the development of recommendation communities.

% \section{Related Work}

\begin{figure}
\centering
\includegraphics[scale=0.19]{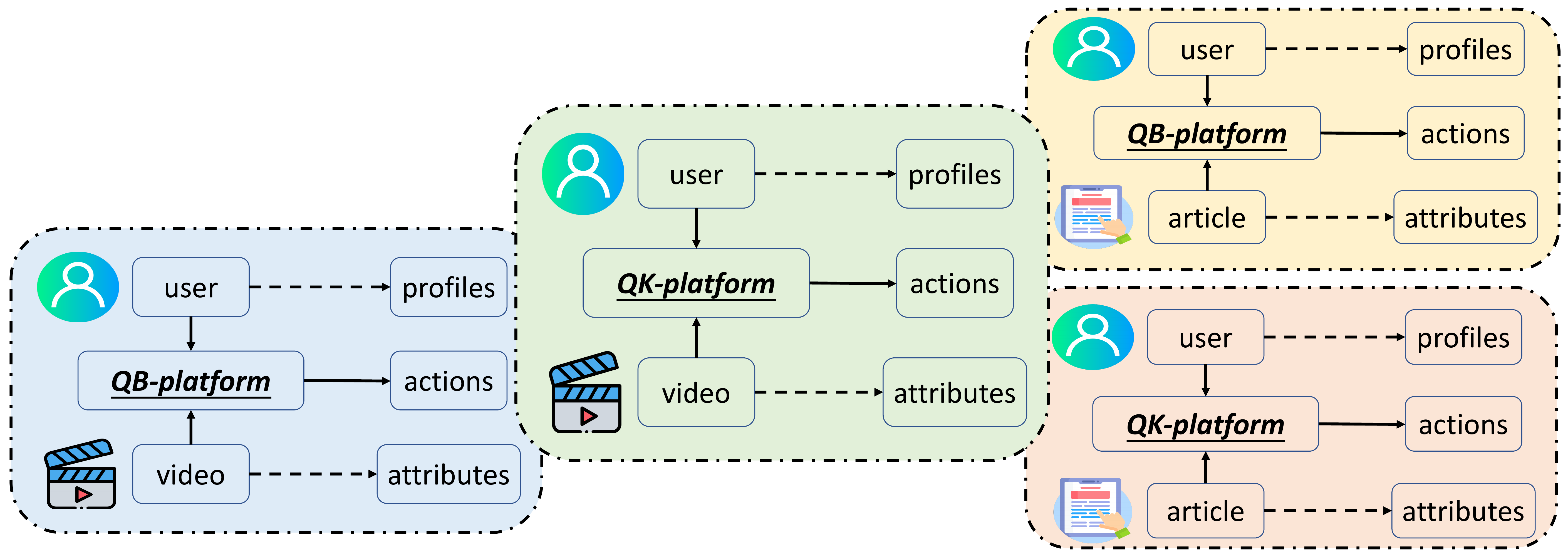}
% \caption{Data sources of Tenrec. \textcolor{red}{Tenrec contains four recommendation scenarios: QK-video, QK-article, QB-video, and QB-article, where there are some percentage of overlapped users and items between every two or three scenarios. }  }
\caption{Data sources of Tenrec. Tenrec contains four recommendation scenarios: QK-video, QK-article, QB-video, and QB-article, where there are some percentage of overlapped users and items between every two or three scenarios.  }
\label{fig:construct}
\end{figure}

% \begin{figure}
% \centering
% \includegraphics[width=6cm]{QB-video.png}

% \caption{ Distribution of clicks per user in QB-video}

% \label{fig:useregistration}
% \end{figure}

\section{Dataset Description}
\label{datades}
% \subsection{Dataset Construction}
Tenrec is a dataset suite developed for multiple recommendation tasks, collected from two different feeds recommendation platforms  of Tencent\footnote{https://www.tencent.com/en-us/}, namely, 
% \textcolor{red}{QQ Browser (QB) and QQ Kandian (QK)}.
% QQ Browser (QB) and QQ Kandian (QK).
QQ BOW (QB) and QQ KAN (QK).\footnote{The names of the two production systems have been anonymized (email us if you want to know their names).}
% a content recommendation platform which we call it QK to distinguish it from the above QB.
%QQ Kandian (QK) 
 An item in  QK/QB  can  either be  a news article or a video. Note that the article and video recommendation models are trained separately with different neural networks and features. Thus, we can think that Tenrec
 is composed of user feedback from four scenarios in total, namely, QK-video, QK-article, QB-video, and QB-article (see Figure~\ref{fig:construct}). We collect user behavior logs from QK/QB from September 17 to December 07, 2021. The procedure is as follows:  we first randomly draw around 5.02 million users from the QK-video database, with the requirement that each user had at least 5 video clicking behaviors; then, we extract their feedback (around 493 million), including both positive feedback (i.e. video click, share, like and follow) and negative feedback (with exposure but no user action); finally, we obtain around 142 million clicks, 10 million likes, 1 million shares and 0.86 million follows, alongside 3.75 million  videos. Besides,
 there are age and gender features for users, and the video type feature for items.
 We perform similar data extraction strategy for  QK-article, QB-video, and QB-article.
%  The major difference is that for
% QK/QB-article, we extract over 20 article features, as described later.
% , such as total numbers of exposures, clicks, likes, the article quality score, and coarse- \& fine-grained  categories.
In this paper, we regard QK-video as the main scene, and other three as the secondary scenes, used for various CDR or TF tasks.
 The dataset statistics are shown in Table~\ref{statistics}.

\begin{table}
  \caption{Data Statistics.  avg \#clicks denotes the average number of clicks per user; \#exposure denotes the number of exposures to users (a.k.a. impressions), including both positive and negative feedback. }
  \label{statistics}
  \centering
  \begin{tabular}{lllll}
    \toprule
    Name     & QK-video  & QK-article  & QB-video & QB-article \\
    \midrule
    \#users & 5,022,750 & 1,325,838 &  34,240 & 24,516\\ 
    \#items & 3,753,436 &  220,122  &130,637 & 7,355 \\ 
    \#click & 142,321,193 &  46,111,728  &1,701,171 & 348,736 \\
    % \#exp & 409,546,120 & 2,413,801 & 33,368,301 & 339,788 \\ 
      \#like & 10,141,195 & 821,888 & 20,687  & / \\ 
    \#share & 1,128,312 & 591,834 & 2,541  & / \\ 
    \#follow & 857,678 & 62,239 & 2,487  & / \\ 
    \#read & / & 44,228,593  &  /& / \\
    % \#switch & / & / & / & 240,961 \\
    \#favorite  & / &  316,627 & / & /\\
         \#exposure & 493,458,970 &  /  & 2,442,299& /\\
    avg \#clicks & 28.34 &34.78  &  49.69 & 14.22 \\ 
    \bottomrule
  \end{tabular}
\end{table}

\begin{figure}[t]
\centering
\subfigure[QK-video]
% {\includegraphics[width=0.3\linewidth]{NeurIPS-22-RS/qkv-ori.pdf}}
{\includegraphics[width=0.3\linewidth]{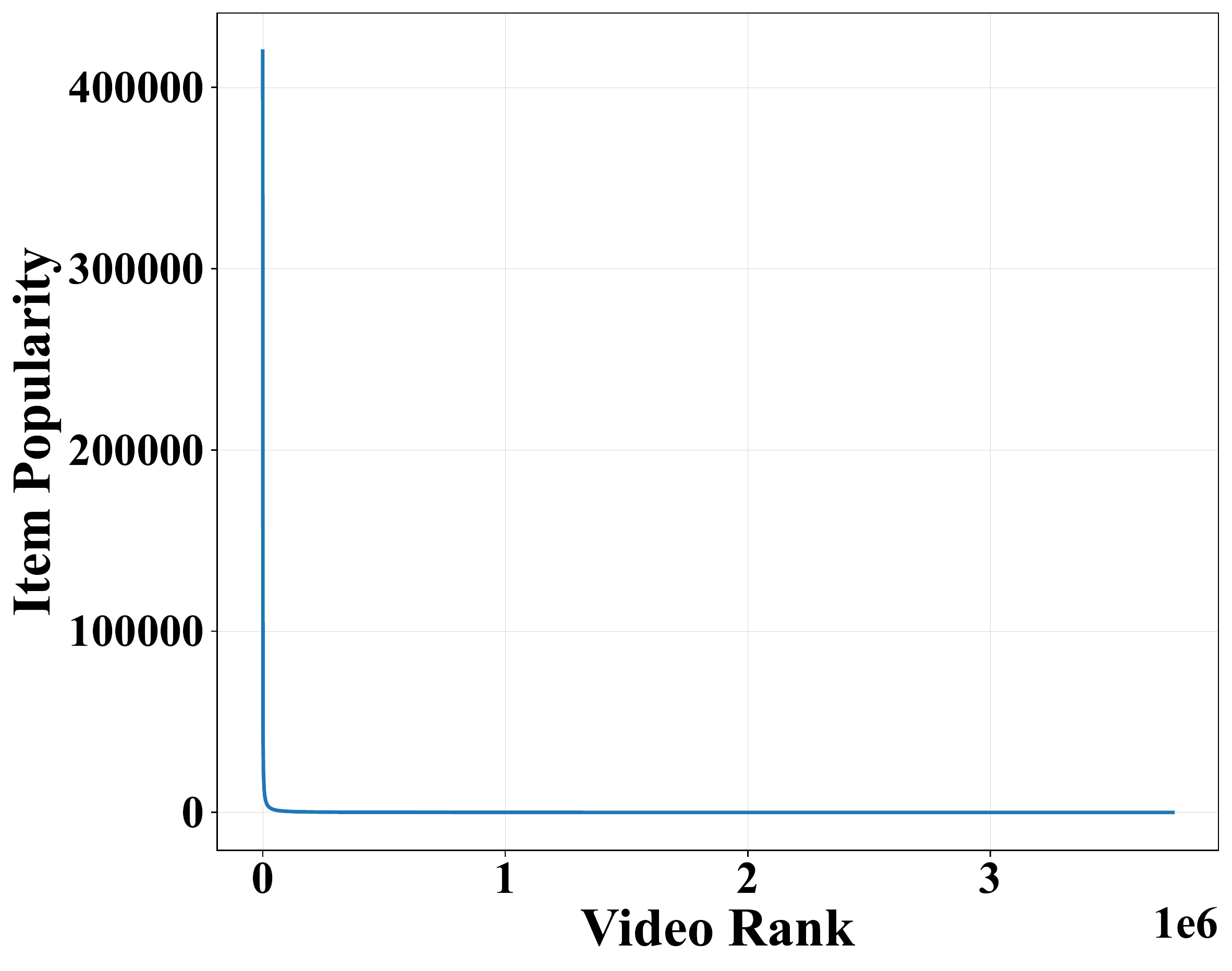}}
\hfill
% \hspace{0.02\linewidth}
\subfigure[QK-video]
% {\includegraphics[width=0.3\linewidth]{NeurIPS-22-RS/qkv-log-log.pdf}}
{\includegraphics[width=0.3\linewidth]{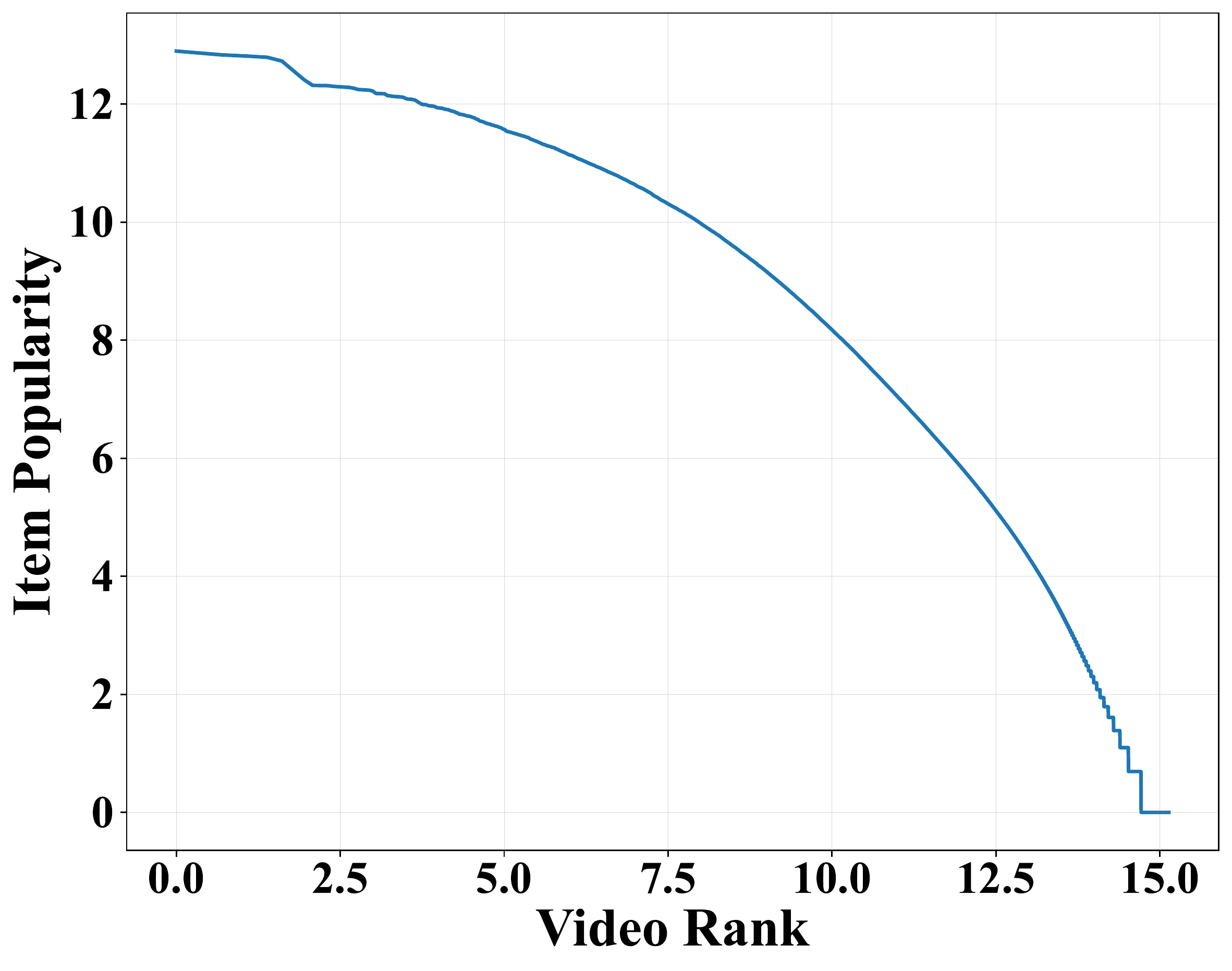}}
\hfill
\subfigure[QK-video]
% {\includegraphics[width=0.3\linewidth]{NeurIPS-22-RS/qkv-user.pdf}}
{\includegraphics[width=0.3\linewidth]{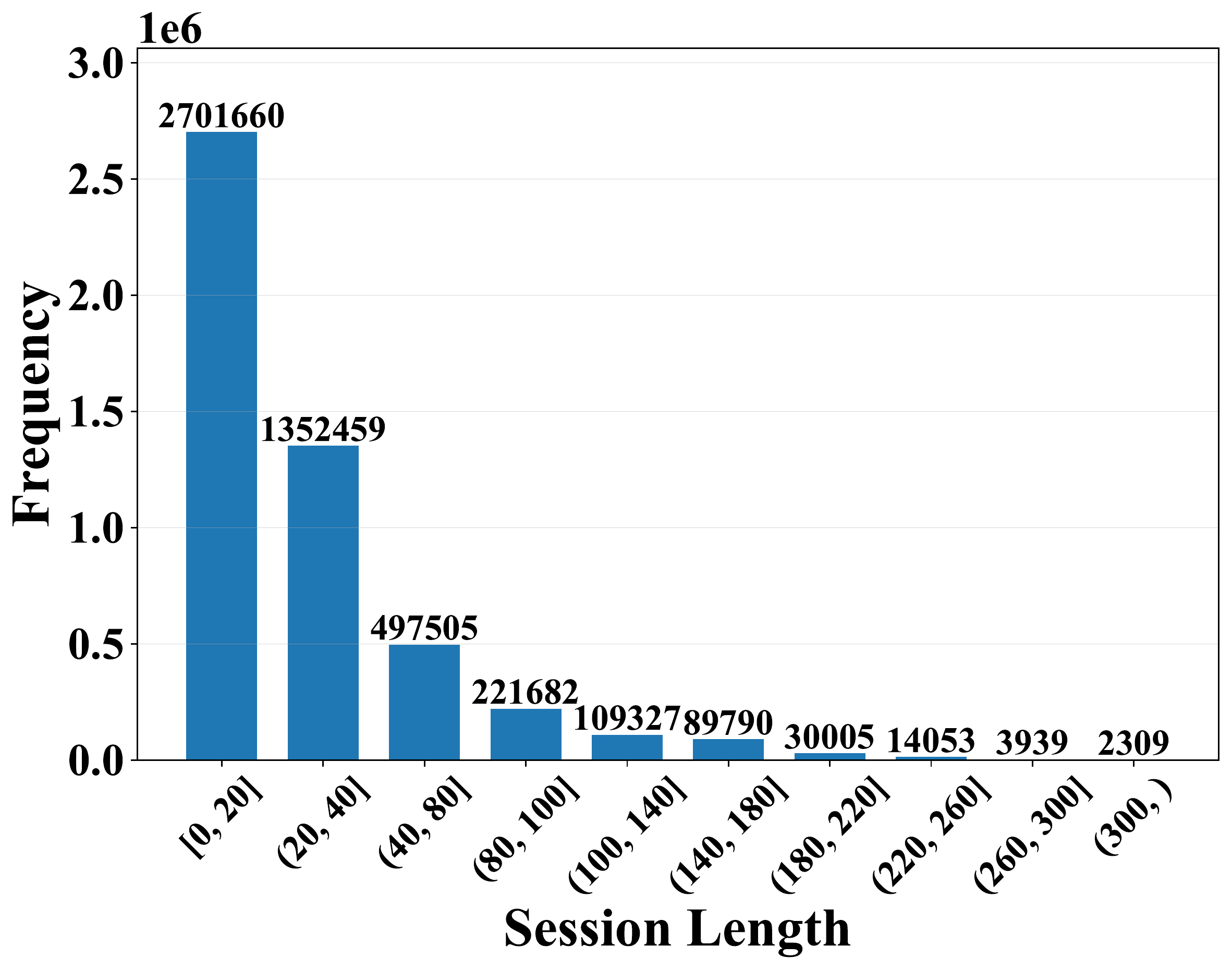}}
% \caption{ Distribution of clicks per user in per dataset}
\caption{ Item distribution.
(a) and  (b) are item popularity plotted in raw and log-log scales; (c) is the item session length distribution.
% x-axis is the ranks in terms of clicking popularity, while y-axis
}
\label{distribution}
\end{figure}

\textbf{Data Distribution.}
Figure\ref{distribution} (a) and (b)  show  the item popularity of QK-video in terms of the clicking behaviors. Clearly, the item popularity follows a  typical long-tail distribution, which has been widely reported in previous recommendation literature~\cite{rendle2014improving}. 
(c) shows the session length distribution, where the number of sessions with length in $[0-20]$ accounts for 53\% of all sessions.
Similar distributions can be observed on the other three datasets, which are thus simply omitted.

% For the clarity purpose. we only show the distribution of 300 most popular items for each dataset. The popularity of remaining items fall into the same distribution.
% It can be clearly seen that item popularity of all the four dataset follows a  typical long-tail distribution, which has been widely observed in previous recommendation literature~\cite{rendle2014improving}. 
% For example, the number of clicks with popularity ranks between [0-20] is 2 times and 10 times larger than those with ranks between [20-40] and  [80-100] respectively on QK-video.
% In particular, the number of clicks with ranks between [0-20]  amounts to 53\% of all clicks for the 300 most popular items.
% % of clicking behaviors in the each dataset, and the number of click behaviors of most users is distributed between 1 and 40. The distribution shows a log-like decay. And there is user overlap between the main scene dataset and other scene datasets, among which there is 9.5\% user overlap between QK-video and QB-video, and 20.23\% user overlap between QK-video and QK-article. The QB-article scene data is less, and only 0.23\% of the users of QK-video and QB-article overlap.

\textbf{Data Overlapping.}
Tenrec contains a portion of overlapped users and items across the four scenarios. Regarding overlapped users, we calculate them between QK-video and QK-article, QB-video, QB-article given that QK-video covers the largest number of users, items and interactions.
Specifically, the number of overlapped users is 268,207 between QK-video and QK-article, 3,261 between QK-video and QB-video, and 58 between QK-video and QB-article. 
 Regarding overlapped items, 78,482 videos  are overlapped  between QK-video and  QB-video.
%  and 129 news articles
 Overlapped users and items can be associated by their unique IDs.
 This property makes Tenrec well-suited for studying the TF and CDR tasks.

\textbf{User Feedback.}
Tenrec is different from existing recommendation datasets,  containing  only one type of user feedback, either implicit feedback or explicit ratings.
% (e.g. clicking or watching) or only user explicit ratings. 
As a result, the degree of user preference in these datasets cannot be well reflected.
% \footnote{While ratings can be dealt with as binary values denoting whether the user has seen, it is not suggested to treat explicit rating scores as the preference degree for the top-$N$ item recommendation task, see~\cite{cremonesi2010performance} for details.} 
As shown in Table~\ref{statistics},  QK-video and QB-video include four types of positive feedback, where clicking behaviors account for the largest number, followed by likes, shares and follows. This finding is  intuitive  since likes, shares and follows often denote much higher preference than clicks. Likewise, QK-article contains two additional types of preference feedback, i.e. the article reading  and favorite behaviors.
% The rich user feedback can be used to evaluate the various MTL tasks and TF-based preference prediction tasks, e.g. transferring click-level preference to like-level preference.
Beyond various positive feedback, Tenrec includes true negative feedback --- i.e. an item is present to a user, but s/he has not clicked it. 
Such negative feedback enables Tenrec to be more suited to CTR prediction, for which most existing datasets involve only positive feedback.

\textbf{Features.}
The format of each instance in QK/QB-video is \{\textit {user ID, item ID, click,  like, share, follow, 
video category, watching times, user gender, user age}\}. Note that the timestamp information has been removed required by Tencent, but we present all interaction behaviors according to the time order.    \textit {click, like, share, follow} are binary values denoting whether the user has such an action. \textit{watching times} is the number of watching behaviors on the video. \textit {user ID, item ID, user gender, user age} have been  desensitized for privacy issues. 
 \textit {User age} has been split into bins, with each bin representing a 10-year period.
% \textit {User age} is divided into 10 classes and each class denotes 10-year-old.
% where user\textit{\_id} is  the anonymous ID of a user, and \textit{item\_id} is the anonymous  ID of an item.  
% %  The behavior characteristics include clicks, exposure, follow, like, share, etc.
% User feedback (e.g. \textit{click, like, share, follow}) is represented by binary values, where 1 means the user has this action and 0 otherwise.
% %  , where 1 means the user has done this behavior, 0 means the user has not done this behavior. 
% \textit{short\_v} means whether the video is a short one.  \textit{play\_times}  is the number of watching times on the video.   Gender and age are user profiles, and timestamp is the time when the clicking action happens.  
% % These three feature are also desensitized for privacy issues.

The format of each instance in QK/QB-article is 
\{\textit {user ID, item ID, click, like, share, follow, read, favorite,
click\_count, like\_count, comment\_count, exposure\_count, 
% commentlike\_count, 
read\_percentage,  category\_second, category\_first, item\_score1, item\_score2, item\_score3,  read\_time}\}.
% gender, age, exposure\_total, click\_total, like\_total, comment\_total, 
% commentlike\_total, read\_completion, quality\_score, media\_score,
% channel\_second, channel\_first, summary\_pic\_score, exposure, exp\_time, read\_time, read, read\_time, share, like, follow, collect}].
% [\textit {user\_id, item\_id,  gender, age, click, like, share, follow, favorite, read, exposure\_total, click\_total, like\_total, comment\_total, commentlike\_total, read\_completion, quality\_score, media\_score, channel\_second, channel\_first, summary\_pic\_score, exp\_time, read\_time, timestamp}].
% % where user\_id is the anonymous ID of a user, and item\_id is the anonymous ID of a item. 
The suffix ``\textit{$*$\_count}'' denotes the total number of $*$ actions per article. 
% \textit{commentlike\_total} denotes the total number of likes of all comments per  article. 
\textit{read\_percentage} denotes how much percentage the user has read the article.
% , with value ranging from 0 to 100. 
\textit{category\_first} and \textit{category\_second} are categories of the article, where ``\textit{\_first}'' is the coarse-grained category (e.g. sports, entertainment, military, etc) and ``\textit{\_second}'' is the fine-grained category (e.g. NBA, World Cup, Kobe, etc.). 
\textit{item\_score1, item\_score2, item\_score3} denote the quality of the item by different scoring system.  \textit{read\_time} is the duration of reading. 

\section{Experimental Evaluation}
In this section, we describe the applications of Tenrec in ten distinct recommendation tasks. We briefly introduce each task and report results of popular or state-of-the-art baselines. We implemented key baselines by referring to the official code, code of DeepCTR\footnote{https://github.com/shenweichen/DeepCTR-Torch} and Recbole\footnote{https://recbole.io}.
% These methods are the classic methods of each scenario, and show their comparison on the dataset.
\subsection{CTR Prediction}
\label{CTR}
CTR prediction is a classical recommendation task where its goal is to predict whether or not a user is going to click a recommended item. We perform this task on the sampling version of the QK-video dataset with 1 million randomly drawn users, referred to as QK-video-1M. More results of the full QK-video dataset are present in Appendix Table 2.

The reason to perform sampling is simply because
% because 
searching hyper-parameters for all baselines of the ten tasks on the original dataset
requires too much compute
% compute 
and training time.
% \textcolor{red}{
% The reason to perform sampling is simply because searching hyper-parameters for all baselines of the ten tasks on the original dataset
% requires too much compute and training time.}
% The reason for sampling is 
% to enable that the dataset could better fit a
% 24G memory graphics processing unit (GPU), which is generally available for academic groups.
We will create a public leaderboard for the original dataset. 
% \yuan{We show  results and make  leaderboards for the original dataset in Appendix XXX.} 

\textbf{DataSet.}
We keep all positive feedback and draw a portion of true negative feedback with positive/negative sampling  ratio of $1:2$.
% We draw negative examples from the true negative feedback for the 1 million users with positive/negative sampling ratio $1:2$,
By doing this, we
obtain in total 1,948,388 items and 86,642,580 interactions with 96.7\% sparsity. Then we split the data into  8:1:1 as the training set, validation set, and testing set.\footnote{We use the  8:1:1 data splitting strategy for all tasks in this paper.} The training example consists of the following features: user ID, item ID, gender, age, video category, and user's past 10 clicked items. We apply  embedding for categorical features. 

\textbf{Baselines and hyper-parameters.} There are a bunch of  deep learning baselines for  CTR prediction.
% However, as reported in ~\cite{zhu2021open}, most of these models perform very  closely with respect to Area Under Curve (AUC)~\cite{rendle2012bpr}. 
Among them, Wide \& Deep ~\cite{cheng2016wide}, DeepFM~\cite{lian2018xdeepfm}, NeuralFM~\cite{he2017neural}, xDeepFM~\cite{lian2018xdeepfm}, Attention FM (AFM)~\cite{xiao2017attentional}, {DCN~\cite{wang2017deep} and DCNv2~\cite{wang2021dcn} are  some of the most well-known and  powerful baselines in  literature~\cite{zhu2021open,wang2021dcn,sun2021fm2}.  
A recent CTR benchmark~\cite{zhu2021open} shows that many recent models (e.g.  InterHAt~\cite{li2020interpretable}, AFN+~\cite{cheng2020adaptive}, and LorentzFM~\cite{xu2020learning}) do not perform significantly better than these popular baselines.
% We run all  baselines using the official code of PyTorch DeepCTR\footnote{https://github.com/shenweichen/DeepCTR-Torch}~\cite{shen2018deepctr}. 

We perform hyper-parameters search on the validation set by evaluating embedding size (denoted as $d$ throughout this paper) in \{16, 32, 64, 128\}, hidden units  (denoted as $f$) in \{64, 128, 256\}, and learning rate  (denoted as $\eta$) in  $\{1e{-3},\ 5e{-4},\ 1e{-4},\ 5e{-5},\ 1e{-5} \}$. 
Finally, we set $\eta$ to $ 5e{-5}$, $d$ to 32 for all methods,  set $f$ to 256 for DeepFM, NFM, Wide \& Deep and xDeepFM, and set the attention factor to 8 for AFM. As for batch size (denoted as $b$), it performs in general slightly better by using a larger one. We set it to 4096 for all models. We find regularization $\lambda$ has no obvious effects on the results, probably  due to much larger training examples, so we set it to $ 5e{-5}$. We set the layer number $h$ to 1 for AFM and 2 for other models according to the optimal results in the validation set.

% Hence, we evaluate Tenrec using these five methods. 

% Click-through rate prediction is the task of predicting the likelihood that something on a website or application (such as an advertisement) will be clicked. Specifically, we predict how likely the user is to click on the video based on the features of each sample (including user attributes and item attributes). To evaluate the performance of LDRC dataset, 5 recommendation algorithms are applied in QB-video dataset. We randomly sample positive and negative samples, their ratio is 1:2. After sampling, we split instances by 8:1:1 for training , validation and test.

% \textbf{WDL}[]: This method proposed by Google, which combines deep neural network and linear model, and widely used in real scenarios.

% \textbf{AFM}[]: The method is a two-stage algorithm AutoFIS to automatically select important low-order and high-order feature interactions in factorization models.

% \textbf{DeepFM}[]: A new neural network model DeepFM that integrates the architectures of FM and deep neural networks (DNN).

% \textbf{NFM}[]: The method uses Bi-Interaction Layer to model second-order feature interactions.

% \textbf{XDeepFM}[]: Different from DeepFM, XDeepFM adds a compressed interaction network(CIN) that learns high-order feature interactions explicitly.
\textbf{Results.}
Table \ref{table:ctr_result} shows the results of different methods on the QK-video-1M dataset in terms of   Area Under Curve (AUC)~\cite{rendle2012bpr}. We observe that in general these CTR models perform very similarly. By  contrast,
NFM performs the best whereas Wide \& Deep performs the worst with around 0.5\% disparity. 
% other methods perform very similarly with AUC from 0.7896 to 0.7928 with around 0.4\% accuracy gap.
% achieving almost the same level of accuracy ($\sim$ 0.7957).
% The results show that the differences among these state-of-the-art models is
% small ($<$ 0.03\%) under the same and fair experimental setting. The finding is  consistent with that in ~\cite{zhu2021open}. 
% Similar observations can be observed in Appendix~\ref{XXX} Table~\ref{XXX}. 

\begin{table}
    \centering
	\begin{minipage}{0.45\linewidth}
		\centering
        \caption{Results for CTR prediction.}
        \label{table:ctr_result}
        \begin{tabular}{lcc}
            \toprule
            Model     & AUC  & Logloss\\
            \midrule
            % Wide \& Deep & 0.7915 \\ 
             Wide \& Deep & 0.7919 & 0.5100\\ 
            DeepFM & 0.7930 &0.5083\\ 
            NFM & 0.7957 & 0.5080\\ 
             xDeepFM &  \textcolor{black}{0.7931} & 0.5081 \\
            AFM & 0.7928 & 0.5090\\
            DCN & 0.7927 & 0.5092\\
            DCNv2 & 0.7932 & 0.5085\\
            \bottomrule
         \end{tabular}
	\end{minipage}
	\hfill
	\begin{minipage}{0.5\linewidth}
		\centering
        \caption{Results for SBR.}
        \label{table:session_result}
        \begin{tabular}{lcc}
            \toprule
            Model     & HR@20 & NDCG@20  \\
            \midrule
            GRU4Rec &0.04882  &0.0192  \\
            NextItNet &0.05112  & 0.0199 \\ 
            SASRec &0.05017  & 0.0194 \\
            BERT4Rec &0.04902  & 0.0185 \\
            \bottomrule
         \end{tabular}
	    \end{minipage}
\end{table}

\subsection{Session-based Recommendation}
\label{sbr}
Session-based recommendation (SBR), a.k.a. sequential recommendation, aims to predicting the next item given a sequence of previous  interacted items in the session, One key feature of SBR is that the interaction orders are explicitly modeled during training, which often yields better top-$N$ results. 
% We refer to Recbole for some code implementation in the SBR tasks.\footnote{https://recbole.io}~\cite{recbole}

\textbf{DataSet.} 
We report baseline results evaluated on QK-video-1M here. 
More results of the full QK-video dataset are present in Appendix Table 3.
Following the common practice~\cite{yuan2019simple},
% Recommendation based on a sequence of events. e.g. next item prediction. It is a kind of top N recommendation.
% Different from traditional recommendation, sequential recommendation needs clear time information. The user's behavior in the data set we provide is arranged in chronological order, and there is no need for the users to arrange the samples by themselves.
% To evaluate the performance of LDRC dataset, 3 recommendation algorithms are applied in QQ-video dataset, and We sorted the items for each user interaction according to time. Due to the large data volume of the dataset, we will use one million users of it for training, evaluation and testing due to gpu device limitations. 
we simply filter out sessions with length shorter than 10. Given that the average session length is 28.34, we set the maximum session lengths to 30. Session length less than 30 will be padded with zero, otherwise only recent 30 interactions are kept. After pre-processing, we obtain 928,562 users, 1,189,341 items and 37,823,609 clicking interactions. We keep the last item in the session for testing, the second to last for validating, and the remaining for training.

\textbf{Baselines and hyper-parameters.}
We verify Tenrec by using four highly cited baselines: RNN-based GRU4Rec~\cite{hidasi2015session,tan2016improved}, CNN-based NextItNet~\cite{yuan2019simple}, self-attention-based SASRec~\cite{kang2018self} and BERT4Rec~\cite{sun2019bert4rec}. In the original paper, these models adopted different loss functions, sampling methods and data augmentations~\cite{tan2016improved}, which are not comparable 
when evaluating   network architectures.
% \footnote{Similarly, we notice that ~\cite{ludewig2021empirical} reported results for over 10 SBS baselines, however, by studying the official code, we find that some baselines were evaluated with different pre-processing strategies, e.g. the session length.} 
Hence,  to perform a rigorous comparison, we apply the standard  autoregressive~\cite{yuan2019simple,yuan2020future} training fashion with
the cross entropy loss and softmax function for GRU4Rec, NextItNet and SASRec --- i.e. only the network architectures of them are different. We train BERT4Rec with the original mask token loss, which is used to compare with SASRec since both of them apply the multi-head self-attention network architecture.
% BERT4Rec is distinct from them, which trains the network by leveraging both the past and future context data. 

% GRU4Rec, NextItNet and SASRec are trained in the autoregressive manner, while BERT4Rec is trained by the masked token loss
% \textbf{BERT4Rec}: It is a Bidirectional Encoder Representations from Transformers for sequential Recommendation (BERT4Rec).

% \textbf{NextItNet}: The network architecture of the proposed model is formed of a stack of holed convolutional layers, which can efficiently increase the receptive fields without relying on the pooling operation.

% \textbf{SASRec}: A sequential model that adopts the self-attention layer to capture the dynamic user interaction sequences.
The hyper-parameters are searched similarly as above. $\eta$ is set to  $5e{-4}$ for GRU4Rec and $1e{-4}$ for all other three models.
% $\{1e{-3},\ 5e{-4},\ 1e{-4},\ 5e{-5},\ 1e{-5} \}$. 
% Bert4rec-0.0001 gru4rec-0.0005 nextitnet-0.0001 sasrec-0.0001
$b$ and $\lambda$ are set to 32 and 0 for all models. $d$, $f$ and $h$ are set to 128, 128 and 16 for NextItNet and BERT4Rec, while they are set to 64, 64 and 8 for GRU4Rec and SASRec. This is simply because SASRec and GRU4Rec produce inferior results with larger $d$, $f$ and $h$ in the validation set. The attention head is set to 4 for SASRec and BERT4Rec, which performs a bit better than 1 and 2. We randomly mask 30\%  items in each  session for BERT4Rec after searching in \{10\%, 20\%, 30\%, 40\%, 50\%\}.

% $d$,  $f$, $\eta$, $b$ and $\lambda$ of all models are set to 128, 128, $1e-4$, 32 and 0, respectively. We empirically find that BERT4Rec performs the best by setting both $d$ and $f$ to 64.
% In addition, the number of hidden layers is set to 8 for BERT4Rec and 16 for all other models according to the results in validation set.

% un BERT4Rec with a XXX\% probability of masking each token during training.
% ed的超参实验是embedding size32、64、128，层数4、8、16，学习率和ctr的设置一样，最后使用0.0001的学习率  ctr的batch size64，session based的batch size32  regular设置都是0

\textbf{Results.} We evaluate all baseline using the standard top-$N$ ranking metrics, i.e. hit ratio (HR)~\cite{yuan2019simple} and normalized discounted cumulative gain (NDCG)~\cite{kang2018self}. $N$ is set to 20. 
% We use NDCG@N (Normalized Discounted Cumulative Gain) metric for model evaluation, N is set to 5 and 20 for comparison, and it is used as an evaluation metric for sorting results, which can better evaluate the accuracy of sorting. We evaluate the prediction accuracy of the last (i.e., next) item of each sequence in the testing set. 
Table \ref{table:session_result} shows the  results of the four baselines. The obervations are as follows: (1) The unidirectional models GRU4Rec, NextItNet and SASRec
offer better results than the bidirectional BERT4Rec on HR@20 and NDCG@20. This is  consistent with many recent works~\cite{zhou2022filter,liu2021c,dallmann2021case}.
(2) With the same training manner,
the three unidirectional models perform similarly --- NextItNet with temporal CNN architecture perform slightly better than SASRec and GRU4Rec. Our results here are different from many previous publications~\cite{sun2019bert4rec,ludewig2021empirical,xie2020contrastive} where the best performed network architecture easily obtains over 50\% improvements over classical baselines.  
% Clearly, SASRec yields the best results than other methods, which indicate (1) the self-attention network is more expressive than CNN and RNN models; (2)
% the BERT-like bidirectional encoder does not outperform the unidirectional autoregressive training fashion (i.e. GRU4Rec, NextItNet and SASRec), which is opposite to the claim in ~\cite{sun2019bert4rec}. 
% the unidirectional autoregressive training fashion (i.e. GRU4Rec, NextItNet and SASRec) is more effective than the BERT-like bidirectional encoder for the SBR task. 

\subsection{Multi-task Learning for Recommendation}
Multi-task learning (MTL) aims to learn two or more tasks simultaneously while maximizing performance on one or all of them. Here, we attempt to model user preference of both clicks and likes rather than only one of them.  We use the same dataset and splitting  strategy as described for CTR prediction. The difference is that we have two output objectives for MTL with one for click and the other for like. Given that Tenrec contains many types of user feedback, one can exploit more objectives to construct more challenging MTL tasks, e.g. three-,  four- or even six-task (i.e. by combing clicks, likes, shares, follows, reads, favorites together) learning using QK-article. 
% Similar to ctr prediction, 
% we not only predict the click probability of the sample, we also predict the user's like probability. To evaluate the performance of LDRC dataset, 2 Multi-Task learning algorithms are applied in QB-video dataset.

\textbf{Baselines and hyper-parameters.}
We evaluate two powerful MTL baselines on Tenrec, namely, MMOE~\cite{ma2018modeling} and ESMM~\cite{ma2018entire}. In addition, we also present the results of single task learning by only optimizing like  or click objective. We set $\eta$, $d$, $f$, $b$ and $h$ to  $1e{-4}$, 32, 128, 4096, 2 respectively after hyper-parameter searching.
% \textbf{ESMM}: It employs a feature representation transfer learning strategy and mainly consists of two sub-networks which adopt the same structure as BASE model. 

% \textbf{MMOE}: The method adapts the Mixture-of-Experts (MoE) structure to multi-task learning by sharing the expert submodels across all tasks, while also having a gating network trained to optimize each task.
\textbf{Results.}  Table \ref{tab:mtl_result} shows the results of four methods on the QK-video-1M dataset in terms of Area Under Curve(AUC)~\cite{rendle2012bpr}. As we see, ESMM performs better than MMOE for both click and like predictions. MMOE does not notably outperform the single objective optimization (SOO). Despite that, MMOE can achieve a good trade-off for two or more objectives simultaneously whereas SOO focuses only on one objective.
% , achieving almost the same level of accuracy(click-AUC:~0.789, like-AUC:~0.903).
% The results show that the differences among these state-of-the-art models is small (< 0.88\%) under the same and fair experimental setting
% The experimental results show that the multi-task model has better generalization performance.

% \begin{table}
%     \centering
% 	\begin{minipage}{0.35\linewidth}
% 		\centering
%         \caption{Results of multi-task Learning.}
%         \label{tab:mtl_result}
%         \begin{tabular}{lll}
%             \toprule
%             Model     & click-AUC  &like-AUC   \\
%             \midrule
%             Only-click & 0.7957 & / \\
%             Only-like & / & 0.893\\ 
%             ESMM & 0.789 & 0.903 \\ 
%             MMOE & 0.785 & 0.895 \\ 
%             \bottomrule
%          \end{tabular}
% 	\end{minipage}
% 	%\qquad
% 	\hfill
% 	\begin{minipage}{0.55\linewidth}
% 		\centering
% 	        \caption{Results of transfer learning with (w/ PT) vs without (w/o PT) pre-training.}
%         \label{tab:tfl_result}
%         \begin{tabular}{lll}
%             \toprule
%             Model     & NDCG@5     & NDCG@20) \\
%             \midrule
%             NextItNet w/o PT & 0.0255 & 0.0473 \\ 
%             NextItNet w/ PT & 0.0268 & 0.0489 \\ 
%             BERT4Rec w/o PT & 0.0220 & 0.0445 \\ 
%             BERT4Rec w/ PT& 0.0239 & 0.0479 \\ 
%             \bottomrule
%          \end{tabular}
% 	    \end{minipage}
% \end{table}

\begin{table}
    \centering
	\begin{minipage}{0.37\linewidth}
		\centering
        \caption{Results for MTL.}
        \label{tab:mtl_result}
        \begin{tabular}{lcc}
            \toprule
            Model     & click-AUC  &like-AUC   \\
            \midrule
            Only-click & 0.7957 & / \\
            Only-like & / & 0.9160\\ 
            ESMM & 0.7940 & 0.9110 \\ 
            MMOE & 0.7900 & 0.9020 \\ 
            PLE & 0.7822 & 0.9103 \\
            \bottomrule
         \end{tabular}
	\end{minipage}
	\hfill
	\begin{minipage}{0.6\linewidth}
		\centering
	        \caption{Results of TF with and without pre-training (PT).}
        \label{tab:tfl_result}
        \begin{tabular}{lcc}
            \toprule
            Model     & HR@20     & NDCG@20 \\
            \midrule
            NextItNet w/ PT &0.12291  & 0.0489 \\
            NextItNet w/o PT &0.11922   & 0.0473 \\
            SASRec w/ PT &0.12612   & 0.0479 \\ 
            SASRec w/o PT &0.11715   & 0.0445 \\ 
            \bottomrule
         \end{tabular}
	    \end{minipage}
\end{table}

\subsection{Transfer Learning for Recommendation}
\label{tfrec}
Transfer learning (TF) --- by first pre-training and then fine-tuning --- has become the de facto  practice in NLP~\cite{devlin2018bert} and CV~\cite{huh2016makes}. However, it remains  unknown what is the best way to perform TF for the recommendation task~\cite{shin2021scaling}. In this section, we simply explore a basic way by first pre-training a SBR model (i.e. NextItNet and SASRec) in the source domain, and then transferring parameters of its hidden layer (i.e. CNN and self-attention) to the same model (with other parameters initialized randomly) in the target domain. We will study other types of transfer learning by considering data overlapping in the following sections.

\textbf{Dataset, baselines, and hyper-parameters.} 
We use the same dataset in the SBR task as the source dataset, and QB-video clicking feedback as the target dataset.
Regarding baseline models, we evaluate the TF effects by using NextItNet and SASRec. Except $\eta$, other hyper-parameters  are set  exactly the same as described in the SBR task.  $\eta$ is set to $1e-4$  and $5e-4$  for SASRec and NextItNet on QB-video.

\textbf{Results.}
Table~\ref{tab:tfl_result} shows the comparison results with \& without pre-training. The key observation is that both NextItNet and SASRec produce better top-N results with pre-training. 
This suggests that parameters of the hidden layers learned from a large training dataset can be used as good initialization for similar recommendation tasks when they have insufficient training data.
\subsection{User Profile Prediction}
\label{userprofile}
User profiles are important features for personalized RS, especially for recommendations of cold/new users. Recently, \cite{yuan2020parameter,chen2021user,cheng2021learning,shin2021one4all} demonstrated that  user profiles could be predicted with high accuracy by modeling their clicking behaviors that are collected from platforms where they have more behaviors. 

\textbf{Dataset, baselines, and hyper-parameters.}
We conduct experiments on the  QK-video-1M dataset. First, we remove instances without user profile features, resulting in
739,737 instances with the gender feature and 741,652 instances with the age feature. 
% only keep examples that contain both user gender and age information,
% \textcolor{red}{select all user whose age and gender are not default, including 739,737 users with gender and 741,652 users with age,} 
We split each dataset into 8:1:1 as the training set, validation set and testing set.
% Both age and gender prediction belong to the typical classification task.
We evaluate five baseline models for this task, namely, the standard DNN model,  PeterRec~\cite{yuan2020parameter} and BERT4Rec
% \footnote{Note that the original paper of BERT4Rec did not investigate its performance for the TF task. Here, we replace NextItNet by BERT4Rec and apply the similar pre-training \& fine-tuning framework.} 
with and without pre-training. The pre-training and fine-tuning framework of PeterRec and BERT4Rec strictly follow~\cite{yuan2020parameter}.
Note that for PeterRec, we use the unidirectional NextItNet as the backbone
% \footnote{In the original paper, they showed that the bidirectional~\cite{yuan2020future} training style performed a bit better for the transfer learning tasks. }
, whereas BERT4Rec is bidirectional. $\eta$ is set to  $1e-4$, $5e-5$, $1e-4$ for DNN, PeterRec and BERT4Rec respectively.
Other hyper-parameters are set the same as in Section~\ref{sbr}.
% PeterRec adopts the parameter-efficient fine-tuning method, whereas BERT adopts the standard full fine-tuning.
%  $\eta$ is set to  $1e-4$  for DNN.
% Except $\eta$ in the fine-tuning stage, all hyper-parameters of PeterRec and BERT4Rec are set  exactly the same as described in the SBR task. 

% the learning rate $\eta$ is set to  $5e-5$ for  NextItNet and PeterRec, $1e-4$ for DNN, BERT4Rec according to the validation set. The embedding size $d$ and layer number $h$ are set to 128 and 16 for DNN. 

% The task is to predict the user's portrait based on the user's click behavior sequence, such as the user's age and gender. And we verify the effect of the pre-trained model on downstream tasks, so we transfer the knowledge learned in pretrain to the training of user's profile, and transfer it to a different task. 3 recommendation algorithms are applied in QB-video dataset.

% \textbf{Peterrec}: Compared with NextItNet, Peterrec has a simple yet very effective grafting network, i.e., model patch, which allows pre-trained weights to remain unaltered and shared for various downstream tasks.

% \textbf{BERT4Rec}: It is a Bidirectional Encoder Representations from
% Transformers for sequential Recommendation (BERT4Rec). 
\textbf{Results.}
Table \ref{tab:upp_result} shows the results of the five baseline models in terms of the standard classification accuracy (ACC).  First, PeterRec and BERT4Rec outperform DNN, indicating that the CNN and self-attention networks are more powerful when modeling user behavior sequences. Second, PeterRec and BERT4Rec with pre-training work better than themselves trained from scratch.

\subsection{Cold-start Recommendation}
\label{coldstart}
Cold start is an important yet unsolved challenge for the recommendation task. A main advantage of Tenrec is that both user overlap and item overlap information is available.
Here, we mainly investigate the cold-user problem by applying transfer learning.  Unlike Section~\ref{tfrec},  both the embedding and the hidden layers can be transferred
for the overlapped users.

% This task is to solve the problem that the user's behavior in a new scene is less, and the user's preferences cannot be effectively learned, so we transfer the knowledge of the existing scene to the cold scene. 
\textbf{Dataset, baselines, and hyper-parameters.}
We treat the QK-video as the source dataset and QK-article as the target dataset. 
% \textcolor{red}{In practice, there are several different cold user recommendation settings. For example,  users tend to have very few clicking interactions in most  advertisement recommender systems, while both warm and cold users could co-exist in other regular recommender systems. Hence, we perform evaluation for several different settings. Since cold users have been removed in our data pre-processing stage, here we simulate a simple cold user scenario by extracting his/her recent 5 interactions from these overlapped users between QK-video and  QK-article. 
% The training, validation and testing set is split into 8:1:1. We have present many additional results for other cold user settings in Table 5 of the Appendix.  } 
In practice, there are several different cold user recommendation settings. For example,  users tend to have very few clicking interactions in most  advertisement recommender systems, while both warm and cold users could co-exist in other regular recommender systems. Hence, we perform evaluation for several different settings. Since cold users have been removed in our data pre-processing stage, here we simulate a simple cold user scenario by extracting his/her recent 5 interactions from these overlapped users between QK-video and  QK-article. 
The training, validation and testing set is split into 8:1:1. We have present many additional results for other cold user settings in Table 5 of the Appendix. 
% We only study item recommendation for users who have interactions less than 5 in the QK-article dataset but have rich interactions in the QK-video dataset. 
We use PeterRec and BERT4Rec as the baseline given their state-of-the-art performance in literature~\cite{yuan2020parameter,yuan2020one,chen2021user,sun2019bert4rec}.
To be specific, we first perform self-supervised pre-training on all  user sequence behaviors in the QK-video  dataset, then fine-tune the model in interactions of these overlapped users between QK-video  and
the QK-article  datasets  to realize the transfer learning. 
More details are given in~\cite{yuan2020parameter}. 
Except $\eta$ in the fine-tuning stage, all hyper-parameters  are set  exactly the same as described in the SBR task. We set $\eta$ to $1e-3$  and $5e-3$  for BERT4Rec and PeterRec respectively during  fine-tuning.
% All hyper-parameters are consistent with Section~\ref{userprofile}.
% We transfer the experience learned in QK-video dataset to the cold scene of QB-video dataset. And capture the first 7 click behaviors of each user as cold-item data.

% \textbf{Peterrec}: Compared with NextItNet, Peterrec has a simple yet very effective grafting network, i.e., model patch, which allows pre-trained weights to remain unaltered and shared for various downstream tasks.

% \textbf{BERT4Rec}: It is a Bidirectional Encoder Representations from Transformers for sequential Recommendation (BERT4Rec). 
\textbf{Results.}
Table~\ref{tab:cd_result} has shown the results of cold-user recommendation. First, we find that both PeterRec and BERT4Rec  yield notable  improvements with pre-training. Second, BERT4Rec with pre-training shows better results than PeterRec.
% in the downstream task than NextItNet with pre-training. 
This is consistent with studies in the  NLP field, where bidirectional encoder enables better transfer learning than the unidirectional encoder. 
% Similar to the transfer learning task, we experiment with the transfer effect of the pre-trained model in the cold start scenario, and the results are shown in Table \ref{tab:cd_result} which have positive effect.

% transferring the pre-trained model to the downstream task of user profile prediction has a positive effect.

% \begin{table}
%     \centering
%     \caption{Results of user profile prediction.}
%     \label{tab:upp_result}
%     \begin{tabular}{lcc}
%         \toprule
%         Model     & Age-ACC  &Gender-ACC   \\
%         \midrule
%         DNN & 0.67875 & 0.88531 \\ 
%          PeterRec w/o PT & 0.68671 & 0.88871 \\
%           PeterRec w/ PT & 0.69712 & 0.90036 \\ 
%          BERT4Rec w/o PT & 0.69151 & 0.89856\\ 
%         BERT4Rec w/ PT & 0.69903 & 0.90082 \\
%         \bottomrule
%      \end{tabular}
% \end{table}

% \begin{table}
% 		\centering
% 		 \caption{Results of cold-start recommendation.}
%         \label{tab:cd_result}
%         \begin{tabular}{lcc}
%             \toprule
%             Model     & HR@20    & NDCG@20  \\
%             \midrule
%             PeterRec w/o PT &0.03571  &  0.0194 \\
%               PeterRec w/ PT& 0.04412  & 0.0221 \\
%             BERT4Rec  w/o PT &0.03555  & 0.0192 \\
%             BERT4Rec w/ PT &0.04963  & 0.0239 \\
%             \bottomrule
%          \end{tabular}
% 	    \end{table}

\begin{table}
    \centering
	\begin{minipage}{0.47 \textwidth}
		\centering
        \caption{Results of user profile prediction.}
        \label{tab:upp_result}
        \setlength{\tabcolsep}{1.5mm}{
        \begin{tabular}{lcc}
            \toprule
            Model     & Age-ACC  &Gender-ACC   \\
            \midrule
            DNN & 0.67875 & 0.88531 \\ 
             PeterRec w/o PT & 0.68671 & 0.88871 \\
              PeterRec w/ PT & 0.69712 & 0.90036 \\ 
             BERT4Rec w/o PT & 0.69151 & 0.89856\\ 
            BERT4Rec w/ PT & 0.69903 & 0.90082 \\
            \bottomrule
         \end{tabular}}
	\end{minipage}
	\hfill
	\begin{minipage}{0.48\textwidth}
		\centering
		 \caption{Results of cold-start recommendation.}
        \label{tab:cd_result}
        \setlength{\tabcolsep}{1.5mm}{
        \begin{tabular}{lcc}
            \toprule
            Model     & HR@20    & NDCG@20  \\
            \midrule
            PeterRec w/o PT &0.03571  &  0.0194 \\
              PeterRec w/ PT& 0.04412  & 0.0221 \\
            BERT4Rec  w/o PT &0.03555  & 0.0192 \\
            BERT4Rec w/ PT &0.04963  & 0.0239 \\
            \bottomrule
         \end{tabular}}
	    \end{minipage}
\end{table}

\subsection{Lifelong User Representation Learning}
When transferring a neural recommendation model from one domain to another, parameters trained for the beginning task tend to be modified to adapt to the new task. As a result, the recommendation model will lose the ability to serve the original task again, termed as catastrophic forgetting~\cite{kirkpatrick2017overcoming}. ~\cite{yuan2020one} proposed the first `one model to serve all' learning paradigm which aims to build a universal user representation (UR) model by using only one backbone network. 
%  However, recommendation datasets involving multiple domains are lacking in literature. 
In this section, we study lifelong learning (LL) by transferring user preference across the four scenarios, i.e. from QK-video to QK-article to QB-video to QB-article.

% An alternative task for future research is to transfer user preference via their feedback, e.g. transferring preference from  clicks to likes to shares to follows.
% from QK-video to QK-article QB-video QB-article

\textbf{Dataset, baselines, and hyper-parameters.} 
For GPU memory issues, we randomly draw 50\% users from QK-video-1M as the dataset for task 1. Then we use QK-article, QB-video and  QB-article for the following tasks.
% We only consider overlapped users between  QK-video-1M  and QK-article, QB-video, QB-article (see details in Section~\ref{datades}).
Given that TF and LL are more favorable to the data scarcity scenario, we process QK-article to remain at most three interactions for each user. Regarding  QB-video and QB-article, we keep their original datasets because the amount of users and clicks is much smaller. 
% Finally, we achieve 79,626, 6,480 and 676 interactions  in QK-article, QB-video and  QB-article respectively. 
% We still use the 8:1:1 splitting strategy to construct training, validation and testing sets. 
Conure~\cite{yuan2020one} is used as the baseline model with NextItNet and SASRec  as the backbone networks. For the comparison purpose, we report results of Conure for task 2, 3, 4 without pre-training (PT) of past tasks. Model-agnostic hyper-parameters are searched similarly as before.   The pruning rates are set to 60\%, 33\%, 25\% for task 1, 2 and 3 respectively.
% Hyper-parameters for task 1 are set the same as the SBR task. 
% For other tasks, they are searched similarly as before with details in Appendix~\ref{}. 

\textbf{Results.}
Table~\ref{tab:life_result} shows the recommendation results with continually learned user representations. It can be clearly seen that  Conure has offered performance improvements on task 2, 3 and 4 due to PT from the past tasks. For example, Conure-NextItNet has increased NDCG@20 from  0.0081 to 0.0095 on task 2,  from 0.0160 to 0.0167 on task 3, and from 0.0902 to 0.1074 on task 4.

\begin{table}[t]
    \centering
    \caption{Results of LL-based cross domain recommendation. NDCG@20 is the evaluation metric.}
    \label{tab:life_result}
    % \scalebox{0.9}{
    \begin{tabular}{lc|c|c|c}
    \toprule
    % \multirow{2}{*}{Model} 
    Model & \multicolumn{1}{c|}{Task1} & \multicolumn{1}{c|}{Task2} & \multicolumn{1}{c|}{Task3} & \multicolumn{1}{c}{Task4} \\
    % & NDCG@20       &NDCG@20       & NDCG@20     & NDCG@20  \\
    \midrule
     Conure-NextItNet w/o PT     & -   &  0.0087  &   0.0162   &  0.0931 \\
     Conure-SASRec    w/o PT   &  -   &  0.0081  &   0.0160   &  0.0902 \\
    Conure-NextItNet    &  0.0177  &  0.0095  &  0.0167  &  0.1074  \\
    Conure-SASRec   &  0.0172     &  0.0086    &   0.0166   &  0.0959 \\
    \bottomrule
\end{tabular}%}
\end{table}

\begin{table}[t]
    \centering
    \caption{Results of model compression. `Cp' is the CpRec framework.  Para. is the number of parameters in millions (M).}
    \label{tab:cp_result}
    \begin{tabular}{lccc}
    \toprule
    Model     & Para. & HR@20  & NDCG@20  \\
    \midrule
    NextItNet & 305M &0.05112  & 0.0199 \\ 
 Cp-NextItNet & 204M &0.05001  & 0.0195 \\ 
    SASRec & 153M &0.05017  & 0.0194 \\ 
    Cp-SASRec & 107M &0.04902  &0.0191 \\  
    \bottomrule
    \end{tabular}
\end{table}

\subsection{Model Compression}
% , large embedding tables often cannot !t in the limited-capacity GPU device
% memory
Model Compression 
% is an actively pursued research area for the machine learning community. It 
enables the deployment of large neural models into  limited-capacity devices, such as GPU and TPU (tensor processing unit). For RS models, the number of parameters in the embedding layer could easily reach   hundreds of millions to billions level. For example,  ~\cite{lian2021persia} recently designed a super large recommendation model with up to 100 trillion parameters. 
% Low-rank factorization
% over the last few years with the goal of deploying state-of-the-art deep networks in low-power and resource limited devices without significant drop in accuracy. 
% Parameter pruning, low-rank factorization and weight quantization are some of the proposed methods to compress the size of deep networks. To compare with session-based recommendation task, we use the same dataset which is QK-video-1M.

\textbf{Dataset, baselines, and hyper-parameters.} We perform parameter compression for the SBR models and use the same dataset as in Section~\ref{sbr}. Despite significant research and practical value, very few work have investigated parameter compression techniques for recommendation task. Here we report results of  the CpRec~\cite{sun2020generic} framework, a state-of-the-art baseline in literature. We instantiate CpRec with NextItNet and 
SASRec as the backbone models. Hyper-parameters are set exactly the same as in Section~\ref{sbr}. 
% Additional hyper-parameters like XX are set as default value in the original paper.
We partition the item set into 3 clusters with partition ratios as $25\%: 50\%: 25\%$ ranked by popularity according to~\cite{sun2020generic}.

% \textbf{Cprec}: The method applies a block-wise adaptive decomposition method to approximate the original large input/output embedding matrices in NextItNet, and proposes layer-wise parameter sharing methods to reduce redundant parameters in the middle layers, which effectively constrains the parameter size as the model grows deeper.

% \textbf{Bert4cp}: Similar to Cprec, we applies block-wise adaptive embedding and layer-wise parameter sharing methods in BERT4Rec.
\textbf{Results.} 
Table \ref{tab:cp_result} shows that CpRec compresses NextItNet and SASRec to two thirds of their original sizes with around 2\% accuracy drop. 
% By comparison, SASRec is relatively more difficult to be compressed since it contains 
% and SASRec to  30\% of the total model size with only marginal accuracy drop.
% \textcolor{red}{and the M of parameter\_num in the ta/ble means millions of parameters}. 
% We adaptively compress the embedding layer by dividing the embedding layer into four adaptive blocks of vocabulary-length quarter, half, three-quarter and full vocabulary length respectively, and then two adjacent blocks in the hidden layer share parameters.
% the memory space compared to the original. We set learning rate to 1e-4 for Cprec.

\subsection{Model Training Speedup}
This task aims to accelerate the training process of very  deep recommendation models. Unlike shallow CTR models, SBR models can be  much deeper.
% Improving training efficiency of large and deep models without sacrificing their performance has attracted much attention in recent years.
Recently,~\cite{wang2020stackrec} revealed that SBR models like NextItNet and SASRec could be deepened up to 100 layers for their best results.\footnote{The authors also released a very large-scale SBR dataset, called Video-6M, which can be used to evaluate very deep (128-layer) RS modes.} To accelerate the training process,
% without overfitting or performance degradation.
they proposed  StackRec, which learns a shallow model first and then copy these shallow layers as top layers of a deep model.  
% Since there are no other suitable baselines, we only report results of StackRec. 
Similarly, we evaluate StackRec by using NextItNet and SASRec as the backbone. Dataset and all-hyper-parameters are kept consistent with Section~\ref{sbr}. 

% We can first simply learn a shallow model with whole data, and then transfer the parameters of the shallow model to the deep model. In the training acceleration task, we accelerate the session-based recommendation, and first train a shallow model in QK-video-1M, and then train a deep model in the same data.

% \textbf{Satckrec}: It is a simple, yet very effective and efficient training framework for deep SR models by iterative layer stacking. Because hidden layers/blocks in a well-trained deep SR model have very similar distributions.

% \textbf{Bert4acc}: Similar to Stackrec, we applies stacking method in BERT4Rec which stacks transformer blocks to speed up model training.

\textbf{Results.} Table \ref{tab:trainspeed} shows the results of training acceleration. Several observations can be made.
(1) StackRec remarkably reduces the training time for both NextItNet and SASRec;
(2) Such training speedup does not lead to a drop in recommendation accuracy. 
% In particular, the stack-SASRec even outperforms its stan version on both NDCCG@5 and HR@20. 
In fact, we even find that Stack64-NextItNet with a 64-layer NextItNet is trained 2$\times$ faster than the standard NextItNet with 16 layers.

% The experiment shows the feasibility of the method, and the acceleration effect of Stackrec is better in comparison. We set the same hyper-parameters as SBR for Stackrec and Bert4acc.

% \begin{table}
%     \centering
% 	\begin{minipage}{0.45\linewidth}
% 		\centering
% 		 \caption{Results of model training speedup. `Stack' denotes the StackRec framework.}
%         \label{tab:acc_result}
% 		  \scalebox{0.85}{
%         \begin{tabular}{llll}
%             \toprule
%             Model   & time &NDCG@5 &NDCG@20  \\
%             \midrule
%             NextItNet & 146900s &0.0103 &0.0199\\ 
%             SASRec & 107200s &0.0101 &0.0194\\ 
%             Stack-NextItNet & 29380s &0.0110 &0.0200\\ 
%             Stack-SASRec & 84544s &0.0100 &0.0196\\ 
%             \bottomrule
%          \end{tabular}}
% 	\end{minipage}
% 	%\qquad
% 	\hfill
% 	\begin{minipage}{0.45\linewidth}
% 		\centering
% 		\caption{Results of model inference speedup. `Skip' means the SkipRec framework}
%         \label{tab:infacc_result}
%     \scalebox{0.85}{
%         \begin{tabular}{llll}
%             \toprule
%             Model     & time & NDCG@5 & NDCG@20  \\
%             \midrule
%             NextItNet & 308s &0.0255 &0.0473\\ 
%             SASRec & 305s &0.0220 &0.0445\\ 
%             Skip-NextItNet & 236s &0.0247 &0.0472\\ 
%             Skip-SASRec & 206s &0.0227 &0.0431\\ 
%             \bottomrule
%          \end{tabular}}
% 	    \end{minipage}
% \end{table}

\subsection{Model Inference Speedup}
As the network goes deeper, a real problem arises: the inference cost increases largely as well, resulting in high latency for online services. ~\cite{chen2021user} showed that users in the recommendation model can be categoried into hard users and easy users, where recommending items to easy users does not pass through the whole network. As a result, authors proposed SkipRec, which adaptively decides which layer is required for which user during model inference phase. 
% similarly as in Section~\ref{}
% We can choose to keep the model blocks or skip it during the inference process at the user's discretion. Similar to the training acceleration task, we perform inference acceleration on session-based recommendation.
% Due to the limitation of GPU memory, 
We verify the effect of model inference acceleration in QB-video.
% \footnote{We notice that the item size in the original paper is much smaller, and thus has no out-of-memory issue.} 
We evaluate SkipRec by assigning NextItNet and SASRec as the backbones. 
Data pre-processing and hyper-parameters are set the same as Section~\ref{sbr}.

% \textbf{Skiprec}: SkipRec is a user-specific depth selection framework where the number of network layers can be selected on a per-user basis. SkipRec enables each user to have their own skipping policies, which is the first personalized depth selection method for the recommendation task.

% \textbf{Bert4infacc}: Similar to Skiprec, we applies a user-specific depth selection framework in BERT4Rec to skip layers which selected by each user.

% For both method, we set up a two-layer policy network to determine which layers of the backbone network can be skipped during inference for .
\textbf{Results.}
Table \ref{tab:infacc_result} shows the effect of SkipRec on QB-video. We see that the skipping policy in SkipRec can  largely speedup the inference time of the SBR models e.g. around 23\% for NextItNet and 32\% for SASRec.  In particular, SkipRec32-NextItNet with a 32-layer NextItNet is still  faster than the original NextItNet with 16 layers.
Moreover, the recommendation accuracy of SkipRec  keeps on par with its original network. 

\begin{table}[t]
	\centering
	 \caption{Results of model training speedup. `Stack' denotes the StackRec framework.}
    \label{tab:trainspeed}
    \begin{tabular}{lccc}
        \toprule
        Model   & Time & HR@20 & NDCG@20  \\
        \midrule
        NextItNet & 66880s &0.05112 & 0.0199 \\ 
        Stack-NextItNet & 24320s &0.05090  & 0.0202 \\ 
        Stack64-NextItNet & 29380s &0.05215  & 0.0200 \\ 
        SASRec & 45040s &0.05017  & 0.0194 \\ 
        Stack-SASRec & 31528s &0.05080  & 0.0196 \\ 
        \bottomrule
     \end{tabular}
\end{table}
\begin{table}[t]
	\centering
	\caption{Results of model inference speedup. `Skip' means the SkipRec framework.}
    \label{tab:infacc_result}
    \begin{tabular}{llll}
        \toprule
        Model     & Time & HR@20 & NDCG@20  \\
        \midrule
        NextItNet & 308s &0.11922  & 0.0473 \\ 
          Skip-NextItNet & 236s &0.12158  & 0.0472 \\
            Skip32-NextItNet & 300s &0.12439  & 0.0484 \\
        SASRec & 305s &0.11715 & 0.0445 \\ 
        Skip-SASRec & 206s &0.11086  & 0.0431 \\ 
        \bottomrule
     \end{tabular}
\end{table}

\section{Conclusions}
We present Tenrec, one of the largest and most versatile recommendation datasets, covering multiple real-world scenarios with various types of user feedback.
% a very large-scale and multipurpose recommendation dataset suite,
% (accessibly hosted on GitHub), 
% which covers four practical recommendation scenarios with various user feedback signals.
To show its  broad  utility, we study it on ten different recommendation tasks and benchmark state-of-the-art neural models in literature. 
 We expose codes, datasets and per-task leaderboards to facilitate research in the  recommendation community, and hope Tenrec becomes a standardized  benchmark to evaluate the progress of these recommendation tasks. Due to 
 % the 
 space limitation, we have
 % has 
 not  explored its full application potential.
 In the future, we plan to (1) investigate Tenrec for more real-world recommendation scenarios, such as cross-domain recommendation with overlapped items~\cite{sheng2021one}, feedback-based transfer learning (e.g. predicting likes and shares based on clicks)~\cite{yuan2020one}, and item recommendation with negative sampling~\cite{yuan2016lambdafm,zhao2014leveraging}; (2)   release
future versions of Tenrec with data that contains item modality information, such as the article title, description, and the raw video contents so as to facilitate multi-modal recommendation~\cite{wang2022transrec,yuan2022}\footnote{We would provide a high-quality multi-modal recommendation dataset with raw images and texts in ~\cite{yuan2022}.}.
%  Due to the space limitation, we has not exhaustively explore its potentials, however, we believe that it can be used to study more practical recommendation scenarios beyond those in this paper.

% In this paper, we propose a large-scale multi-scene dataset and corresponding multi-scene benchmark, which is suitable for more recommendation scenarios. The dataset contains more than 6,200,000 users, 3,800,000 items and hundreds of millions of interactions(click, exp, share, follow, like, etc.). In particular, it also includes multiple scenarios that can be used for transfer learning, cold start and lifelong learning, etc. We experiment with up to ten recommender system scenarios on the dataset, including CTR prediction, session-based recommendation, multi-task learning for recommendation, transfer learning for recommendation, life-long learning for recommendation, recommendation model compression, recommendation model training speedup, recommendation model inference speedup, user profile prediction and cold-start recommendation, which covers the current mainstream research directions of the recommendation community. Not only that, our dataset also contains a lot of rich information to be mined, but due to the space problem we do not expand the relevant content in the text, we believe that this information can be applied to other fields of recommendation system We hope that the dataset and Benchmark can promote the research and development of the recommendation system community.
% \bibliography{rs_benchmark.bib}

\section*{Acknowledgement}
This work is supported by the Research Center for Industries of the Future (No. WU2022C030) and Shenzhen Basic Research Foundation (No. JCYJ20210324115614039 and No. JCYJ20200109113441941).

{
\small
\bibliographystyle{plain}
\bibliography{rs_benchmark.bib}
}

\newpage
\appendix

% \nolinenumbers
\textbf{\Large{Appendix}}
% \section{Appendix}
% \section{Dataset and Code Link}
% Tenrec data link:

% \url{https://drive.google.com/file/d/1R1JhdT9CHzT3qBJODz09pVpHMzShcQ7a/view?usp=sharing}

% Datasets for the ten reported tasks:

% \url{https://drive.google.com/file/d/1ss7QYHvQtfzOF1E31VrWR-_XkNHz-Jfd/view?usp=sharing}

% Tenrec data link \& Baseline code link:

% \url{https://github.com/yuangh-x/2022-NIPS-Tenrec}.

\section{Dataset Comparison}
% To show the difference between Tenrec and other recommendation dataset, we propose Table\ref{tab:dataset_comparison}.
We show the difference between Tenrec and other popular recommendation datasets in Table\ref{tab:dataset_comparison}.
First,
% As shown in the table, Among them, the recommendation dataset with the largest interaction scale (ZhihuRec) contains nearly 10 million interactions. Compared with it, Tenrec has nearly twice as many interactions as zhihurec.
most datasets contain only a single scenario. Without overlapped users and items, it is difficult to develop and evaluate transfer learning recommendation methods. In addition, Tenrec contains very rich positive user feedback, which can be used to evaluate the multi-task learning and preference-level transfer learning tasks.  Third, compared with most recommendation datasets,
% In contrast, our dataset has overlaps of users and items, which can make our datasets suitable for tasks such as transfer learning, cold start, and lifelong learning. 
% It is worth mentioning that our user feedback is very rich, including click, like, share and follow, etc. Therefore, Tenrec can be used in multi-tasks learning scenarios, that is, it can simultaneously predict multiple user behaviors (click, like, share, and follow, etc.). 
% compared with the most datasets of recommendation systems, 
Tenrec has true negative examples, which can be used to evaluate more realistic CTR prediction task. 

It is worth mentioning that
\textbf{the multiple domains in Amazon are defined differently from Tenrec}. In our Tenrec, items of 
 different domains are either from different recommender systems or recommended by completely different algorithms. However,  domains in Amazon are  divided simply based on  their item categories. 
%  In other words, they are more like from the same domain.
 It is unknown whether items of different categories are recommended by the same or different algorithms.  \textbf{It is  not suitable to be used for the  cross domain recommendation tasks if items are recommended by the same  model and from the same platform.}
 In fact, our Tenrec-QKA also includes many different article categories. 
%  In view of this, our Tenrec is more useful for various cross domain recommendation tasks and transfer learning tasks.
%  In view of this, our Tenrec is still more useful, which contains more user feedback and 
%  To summarize, \textbf{Tenrec is by far the only dataset that contain overlapped users and items from completely different recommender systems}.

\begin{table}[t]
    \centering
    \caption{Statistics of popular recommendation datasets. K and M are short for thousand and million respectively. Type$\_$Feeds denotes the number of types of positive user feedback. True$\_$neg denotes whether it includes true negative feedback.
    We only show the statistics of the QK-video (QKV) and QK-article  (QKA) in this table. \#Interactions in Tenrec denotes the clicking behavior.
    }
    \label{tab:dataset_comparison}
    \begin{tabular}{lcccccc}
    \toprule
     Dataset &  Domains & Type\_Feeds &  \#Users  & \#Interactions & True\_neg \\
    \midrule 
    Movielens-20M\footnote{\url{https://grouplens.org/datasets/movielens/20m/}} & Single & 1 &138K   &20M  &\ding{55}  \\
    Amazon\footnote{\url{https://nijianmo.github.io/amazon/index.html}} & Multiple & 2 &/  &233M  &\ding{55}  \\
    Yelp\footnote{\url{https://www.yelp.com/dataset}} & Single & 1 & 1.9M & 8M &\ding{55}  \\
    YOOCHOOSE\footnote{\url{https://www.kaggle.com/datasets/chadgostopp/recsys-challenge-2015?select=dataset-README.txt}}  & Single & 2 & 9.2M & 34M &\ding{55}  \\

   Taobao: User-Behavior\footnote{\url{https://tianchi.aliyun.com/dataset/dataDetail?dataId=649}} & Single & 4 & 987K & 100M &\ding{55} \\
    Ali\_Display\_Ad\_Click\footnote{\url{https://tianchi.aliyun.com/dataset/dataDetail?dataId=56}} & Single & 4 & 1.1M & 26M &\ding{51} \\
    
    TMALL\footnote{\url{https://tianchi.aliyun.com/dataset/dataDetail?dataId=53}} & Single & 2 & 963K  & 44M &\ding{51}   \\
    Yahoo! Music\footnote{\url{https://webscope.sandbox.yahoo.com/catalog.php?datatype=r}} & Single & 1 & 1.9M & 11M  &\ding{55}  \\
    Book-Crossing\footnote{\url{https://www.kaggle.com/datasets/somnambwl/bookcrossing-dataset}} & Single & 1 & 92K & 1.0M  &\ding{55}  \\
    MIND\footnote{\url{https://msnews.github.io}} &Single  &1  &1.0M  &24M  &\ding{55}  \\
    KuaiRec\footnote{\url{https://chongminggao.github.io/KuaiRec/}} &Single  &1  &7K  &12M  &\ding{51}  \\
    ZhihuRec\footnote{\url{https://github.com/THUIR/ZhihuRec-Dataset}} &Single  &1  &798K  &99M  &\ding{51}  \\
    \midrule
    Tenrec-QKV &Multiple  &4  &5.0 M  &142M  &\ding{51} \\
     Tenrec-QKA &Multiple  &6  &1.3 M  &46M  &\ding{55} \\
    \bottomrule 
    \end{tabular}
\end{table}

\section{Supplementary experiment}
In the main body, we only report results with randomly sampled 1 million users, here we show results of the CTR (Table~\ref{table:ctr_5m_result}) prediction and SBR (Table~\ref{table:session_5m_result}) tasks with 5 million users on the full QK-video dataset, following the same experimental setup.
For each task, we  report several top ranked baselines in the main body.

In addition to the above experiments, 
we supplement the experiments of shared historical embedding (i.e., all interacted items share the same embedding) in the CTR prediction task. As show in Table~\ref{table:ctr_share_result},
we could make two observations: (1) CTR models with the shared historical embedding in general slightly underperform models with separate historical embedding (SHE); (2) CTR models with shared historical embedding  show similar accuracy rank as previously reported in Table 2 with SHE.

We also add another experiments with more cold-start settings. To be specific, we notice in some practical recommendation scenarios where both cold and warm users co-exist.
To create such a scenario, we first draw overlapped users between QK-video and QK-article. Then we randomly sample n\% users (e.g., $n=30, 70, 100$ ) and then select the latest $k$ interacted items of them where $k$ is a random integer from 1 to 5, ensuring that these users are cold. The behaviors of the remaining warm users are kept the same. For training, we use all behaviors of these warm users and 50\% behaviors from the cold users. For evaluation, we only evaluate the predictive accuracy for these cold users with 25\% interactions for validation and  25\%  for testing. 
% The behaviors of these warm users are only used for training.
 Results are reported in Table~\ref{table:cold_start_rate_result}.
%  Similar observations can be made as reported in \textcolor{red}{the main body} Table~7.

% and the experiments of coexistence of cold and hot users includes scenarios where the proportion of cold users is 0.3, 0.7 and 1, respectively, in Cold-start task. here we show results of the CTR (Table~\ref{table:ctr_share_result}) prediction with share historical embedding and Cold-start (Table~\ref{table:cold_start_rate_result}) tasks at different percentages of cold users. We set the same experimental hyper-parameters as in the main body, except the scenarios with a cold user ratio of 0.3 and a cold user ratio of 0.7 use a learning rate of 5e-5 in Cold-start task. Table~\ref{table:ctr_share_result} shows that the experimental results of shared embedding are slightly worse than the experimental results of separating historical behaviors by time dimension.}

% Collaborative Filtering is a classical recommendation task where its goal is to address some of the limitations of content-based filtering, collaborative filtering uses similarities between users and content to provide recommendations. 
Here, we report baseline results for the standard top-N item recommendation task on QB-video.
% We report baseline results evaluated on QB-video. 
We filter out users with session length shorter than 10. Then, we split interactions of each user into 8:1:1 as the training set, validation set, and testing set. We evaluate four popular baseline: MF~\cite{koren2009matrix}, NCF~\cite{he2017neural_ncf}, NGCF~\cite{wang2019neural}, LightGCN~\cite{he2020lightgcn} to verify Tenrec. 
The hyper-parameters are searched similarly as before. Learning rate is set to  $5e{-4}$ for NGCF, $1e{-6}$ for NCF, $5e{-3}$ for MF and LightGCN. Batch size is set to 4096 for all models. The embedding size is set to 128 for all models. Then the layer number is set to 2 for NCF, NGCF and LightGCN. 
We show results using two types of negative samplers: random sampler and popularity sampler used in word2vec~\cite{mikolov2013distributed} with power set to 0.75. The number of negative examples is set to 4 for each user.
All results are reported in Table~\ref{tab:cf_ran} and Table~\ref{tab:cf_pop}. It is worth mentioning that more powerful negative samplers could easily lead to better recommendation accuracy than the random and popularity samplers, e.g. the two dynamic samplers used in LambdaFM~\cite{yuan2016lambdafm}. In other words, if you want to compare network architectures, you should ensure that all other settings (loss function, sampling ratio and distribution) are kept the same for comparison.
% By contrast, LightGCN and NGCF perform better result than MF and NCF. Then two negative sampling methods perform similarly.

For other tasks, we would  create per-task leaderboards for the full dataset  version and the 1 million user version.

\begin{table}[t]
    \centering
	\begin{minipage}{0.4\linewidth}
		\centering
        \caption{Results for CTR prediction.}
        \label{table:ctr_5m_result}
        \begin{tabular}{lcc}
            \toprule
            Model     & AUC  &Logloss\\
            \midrule
            Wide \& Deep & 0.8234 &0.4745\\ 
            DeepFM & 0.8235 &0.4741\\ 
            NFM & 0.8231 &0.4750\\ 
            xDeepFM & 0.8235 &0.4740\\
            AFM & 0.8226 &0.4757\\
            \bottomrule
         \end{tabular}
	\end{minipage}
	\hfill
	\begin{minipage}{0.5\linewidth}
		\centering
        \caption{Results for SBR.}
        \label{table:session_5m_result}
        \begin{tabular}{lcc}
            \toprule
            Model     & HR@20 & NDCG@20  \\
            \midrule
            % GRU4Rec &0.04882  &0.0192  \\
            NextItNet &0.05490  & 0.0214 \\ 
            SASRec &0.05164  & 0.0201 \\
            BERT4Rec &0.05027  & 0.0191 \\
            \bottomrule
         \end{tabular}
	    \end{minipage}
\end{table}

\begin{table}[h]
    \centering
	\begin{minipage}{0.4\linewidth}
		\centering
        \caption{Results for CTR prediction with shared historical embeddings on  QK-video-1M.}
        \label{table:ctr_share_result}
        \begin{tabular}{lcc}
            \toprule
            Model     & AUC & Logloss \\
            \midrule
            
            Wide \& Deep & 0.7910 &0.5111 \\ 
            % DeepFM & 0.8244 \\ 
            DeepFM & 0.7920 & 0.5105\\ 
            NFM & 0.7924 &0.5094\\ 
            xDeepFM & 0.7922 & 0.5092\\
            AFM & 0.7921 & 0.5097\\ 
            DCN & 0.7911 & 0.5100\\
            DCNv2 & 0.7922 & 0.5097\\
            DIN~\cite{zhou2018deep} & 0.7910 & 0.5110\\
            DIEN~\cite{zhou2019deep} & 0.7918 & 0.5108\\
            \bottomrule
         \end{tabular}
	\end{minipage}
	\hfill
	\begin{minipage}{0.5\linewidth}
		\centering
        \caption{Results of top-n item recommendation with the random negative sampler.}
        \label{tab:cf_ran}
        \begin{tabular}{lcc}
            \toprule
            Model   & Recall@20 & NDCG@20  \\
            \midrule
            MF &0.0838 &0.0437  \\ 
            NCF &0.0764  &0.0403  \\ 
            LightGCN &0.1065  &0.0542  \\ 
            NGCF &0.0878  &0.0455 \\ 
            \bottomrule
         \end{tabular}
        \vspace{1em}
         \caption{Results of top-n item recommendation  with the popularity negative sampler.}
        \label{tab:cf_pop}
        \begin{tabular}{lcc}
            \toprule
            Model   & Recall@20 & NDCG@20  \\
            \midrule
            MF &0.0927 &0.0467 \\ 
            NCF &0.0757  &0.0405 \\ 
            LightGCN &0.1211  &0.0617 \\ 
            NGCF &0.0948  &0.0476 \\ 
            \bottomrule
         \end{tabular}
        % \caption{Results for cold start prediction with different percentages of cold users. E.g. $cold\_rate \ 0.3$ means that the percentage of cold users in the training data accounts for 30\% of the overlapped users (including both cold and warm users).}
        % \label{table:cold_start_rate_result}
        % \begin{tabular}{llc}
        %     \toprule
        %     Rate &Model & NDCG@20  \\
        %     \midrule
        %     \multirow{5}*{cold\_rate 0.3} &PeterRec w/o PT &0.0112 \\ 
        %     &PeterRec w/ PT &0.0133 \\
        %     &BERT4Rec w/o PT &0.0115 \\
        %     &BERT4Rec w/ PT &0.0137 \\ \hline
            
        %     \multirow{5}*{cold\_rate 0.7} &PeterRec w/o PT &0.0123  \\ 
        %     &PeterRec w/ PT &0.0132 \\
        %     &BERT4Rec w/o PT &0.0119 \\
        %     &BERT4Rec w/ PT &0.0134 \\ \hline
            
        %     \multirow{5}*{cold\_rate 1} &PeterRec w/o PT &0.0158   \\ 
        %     &PeterRec w/ PT &0.0165 \\
        %     &BERT4Rec w/o PT &0.0153 \\
        %     &BERT4Rec w/ PT &0.0166 \\ 
        %     \bottomrule
        %  \end{tabular}
	    \end{minipage}
\end{table}

\begin{table}[t]
    \centering
    \caption{Results for cold start prediction with different percentages of cold users. E.g. $cold\_rate \ 0.3$ means that the percentage of cold users in the training data accounts for 30\% of the overlapped users (including both cold and warm users). NDCG@20 is the evaluation metric.}
        \label{table:cold_start_rate_result}
        \begin{tabular}{lccc}
            \toprule
            Model & cold\_rate 0.3 & cold\_rate 0.7 & cold\_rate 1  \\
            \midrule
            PeterRec w/o PT &0.0112 &0.0123 &0.0158 \\ 
            PeterRec w/ PT &0.0133 &0.0132 &0.0165 \\
            BERT4Rec w/o PT &0.0115 &0.0119 &0.0153 \\
            BERT4Rec w/ PT &0.0137 &0.0134 &0.0166 \\ 
            
            \bottomrule
         \end{tabular}
\end{table}

% \begin{table}
%     \centering
% 	\begin{minipage}{0.4\linewidth}
% 		\centering
%         \caption{Results of top-n item recommendation with the random negative sampler.}
%         \label{tab:cf_ran}
%         \begin{tabular}{lcc}
%             \toprule
%             Model   & Recall@20 & NDCG@20  \\
%             \midrule
%             MF &0.0838 &0.0437  \\ 
%             NCF &0.0764  &0.0403  \\ 
%             LightGCN &0.1065  &0.0542  \\ 
%             NGCF &0.0878  &0.0455 \\ 
%             \bottomrule
%          \end{tabular}
% 	\end{minipage}
% 	\hfill
% 	\begin{minipage}{0.5\linewidth}
% 		\centering
%         \caption{Results of top-n item recommendation  with the popular negative sampler.}
%         \label{tab:cf_pop}
%         \begin{tabular}{lcc}
%             \toprule
%             Model   & Recall@20 & NDCG@20  \\
%             \midrule
%             MF &0.0927 &0.0603 \\ 
%             NCF &0.0757  &0.0405 \\ 
%             LightGCN &0.1211  &0.0792 \\ 
%             NGCF &0.0948  &0.0476 \\ 
%             \bottomrule
%          \end{tabular}
% 	    \end{minipage}
% \end{table}

% \begin{table}[t]
% 	\centering
% 	 \caption{Results of Collaborative Filtering.}
%     \label{tab:cf}
%     \begin{tabular}{lcc}
%         \toprule
%         Model   & Recall@20 & NDCG@20  \\
%         \midrule
%         MF &0.05112 & 0.0199 \\ 
%         NCF &0.05090  & 0.0202 \\ 
%         LightGCN &0.05215  & 0.0200 \\ 
%         NGCF &0.05017  & 0.0194 \\ 
%         \bottomrule
%      \end{tabular}
% \end{table}

\section{General Datasheet of Dataset}

\subsection{Motivation}

\textbf{For what purpose was the dataset created?} Was there a specific task in mind? Was there a specific gap that needed to be filled? Please provide a description.

To foster diverse recommendation research, we propose Tenrec, a large-scale and multipurpose real-world dataset. Compared with existing public datasets, Tenrec has several merits: (1) it consists of overlapped users/items from four different real-world recommendation scenarios, which can be used to study the cross-domain recommendation (CDR) and transfer learning (TF) methods; (2) it contains multiple types of positive user feedback (e.g. clicks, likes, shares, follows, reads and favorites), which can be leveraged to study the multi-task learning (MTL) problem; (3) it has both positive user feedback and true negative feedback, which can be used to study more practical  click-through rate (CTR)  prediction scenario; (4) it has additional user and item features beyond the identity information (i.e. user IDs and item IDs), which can be used for context/content-based recommendations. 

\textbf{Who created the dataset (e.g., which team, research group) and on
behalf of which entity (e.g., company, institution, organization)?}

The dataset was created by Guanghu Yuan and Beibei Kong who were an intern and employee respectively at Tencent. 
% University of Science and Technology of China, Fajie Yuan at Westlake University and Beibei Kong at Tencent.

\textbf{Who funded the creation of the dataset?} If there is an associated grant,
please provide the name of the grantor and the grant name and number.

No.

\subsection{Composition}
\textbf{What do the instances that comprise the dataset represent (e.g., documents, photos, people, countries)?} Are there multiple types of instances (e.g., movies, users, and ratings; people and interactions between them; nodes and edges)? Please provide a description.

The instances are user feedback collected from two different feeds recommendation platforms of Tencent, including both positive feedback (i.e. video click, share, like and follow) and negative feedback (with exposure but no user action). 

\textbf{How many instances are there in total (of each type, if appropriate)?} 

There are 493,458,970 instances in QK(QQ Kandian) Video Datset, 11,722,249 instances in QB(QQ Browser) Video Datset,  46,111,728 instances in QK(QQ Kandian) Article Dataset, and 348,736 instances samples in QB(QQ Browser) Article Dataset, where each sample is user-item interactions.

\textbf{Does the dataset contain all possible instances or is it a sample (not
necessarily random) of instances from a larger set?} If the dataset is
a sample, then what is the larger set? Is the sample representative of the
larger set (e.g., geographic coverage)? If so, please describe how this
representativeness was validated/verified. If it is not representative of the
larger set, please describe why not (e.g., to cover a more diverse range of
instances, because instances were withheld or unavailable).

The dataset is a sample of instances. we randomly draw instances from two different feeds recommendation platforms of Tencent, with the requirement that each user had at least 5 video clicking behaviors. No tests were run to
determine representativeness.

\textbf{What data does each instance consist of?} “Raw” data (e.g., unprocessed text or images)or features? In either case, please provide a description.

The format of each instance in QK/QB-video is \{\textit {user ID, item ID, click,  like, share, follow, 
video category, watching times, user gender, user age, timestamp}\}.   \textit {click, like, share, follow} are binary values denoting whether the user has such an action. \textit{watching times} is the number of watching behaviors on the video. \textit {user ID, item ID, user gender, user age and timestamp} have been  desensitized for privacy issues. 
 \textit {User age} has been split into bins, with each bin representing a 10-year period.
 
 The format of each instance in QK/QB-article is 
\{\textit {user ID, item ID, click, like, share, follow, read, favorite,
click\_count, like\_count, comment\_count, exposure\_count, 
read\_percentage,  category\_second, category\_first, item\_score1, item\_score2, item\_score3,  read\_time, timestamp}\}.
The suffix ``\textit{$*$\_count}'' denotes the total number of $*$ actions per article. 
\textit{read\_percentage} denotes how much percentage the user has read the article, with value ranging from 0 to 100. \textit{category\_first} and \textit{category\_second} are categories of the article, where ``\textit{\_first}'' is the coarse-grained category (e.g. sports, entertainment, military, etc) and ``\textit{\_second}'' is the fine-grained category (e.g. NBA, World Cup, Kobe, etc.). 
\textit{item\_score1, item\_score2, item\_score3} denote the quality of the item by different scoring system.  \textit{read\_time} is the duration of reading. 

\textbf{Is there a label or target associated with each instance?} If so, please
provide a description.

The labels are binary values denoting whether the user has such an action, or user profile.

\textbf{Is any information missing from individual instances?} If so, please
provide a description, explaining why this information is missing (e.g., because it was unavailable). This does not include intentionally removed
information, but might include, e.g., redacted text.

A small percentage of instances lack video category, user age and user gender. The corresponding information is missing in the real system.

% \textcolor{blue}{\textbf{Are relationships between individual instances made explicit (e.g.,
% users’ movie ratings, social network links)?} If so, please describe how these relationships are made explicit}

\textbf{Are there recommended data splits (e.g., training, development/validation, testing)?} If so, please provide a description of these
splits, explaining the rationale behind them.

% In session-based task, We keep the last item in the
% session for testing, the second to last for validating, and the remaining for training. In CTR task, 
 we split the data into 8:1:1 as the training set, validation set, and testing set following some common practice.
%  We divide the dataset according to commonly used data processing

\textbf{Are there any errors, sources of noise, or redundancies in the
dataset? }If so, please provide a description.

No

\textbf{Is the dataset self-contained, or does it link to or otherwise rely on
external resources (e.g., websites, tweets, other datasets)?} If it links to or relies on external resources, a) are there guarantees that they will exist, and remain constant, over time; b) are there official archival versions of the complete dataset (i.e., including the external resources as they existed
at the time the dataset was created); c) are there any restrictions (e.g., licenses, fees) associated with any of the external resources that might apply to a dataset consumer? Please provide descriptions of all external resources and any restrictions associated with them, as well as links or other access points, as appropriate.

The dataset is entirely self-contained.

\textbf{Does the dataset contain data that might be considered confidential
(e.g., data that is protected by legal privilege or by doctor–patient confidentiality, data that includes the content of individuals’ nonpublic communications)? }If so, please provide a description.

No

\textbf{Does the dataset contain data that, if viewed directly, might be offensive, insulting, threatening, or might otherwise cause anxiety? }If so,
please describe why.

No

\textbf{Does the dataset identify any subpopulations (e.g., by age, gender)?
}If so, please describe how these subpopulations are identified and provide
a description of their respective distributions within the dataset.

No

\textbf{Is it possible to identify individuals (i.e., one or more natural persons), either directly or indirectly (i.e., in combination with other
data) from the dataset? }If so, please describe how.

It is impossible to identify individuals from the dataset information

\textbf{Does the dataset contain data that might be considered sensitive in
any way (e.g., data that reveals race or ethnic origins, sexual orientations, religious beliefs, political opinions or union memberships, or
locations; financial or health data; biometric or genetic data; forms of
government identification, such as social security numbers; criminal
history)? }If so, please provide a description

No

\subsection{Collection Process}

\textbf{How was the data associated with each instance acquired? }Was the
data directly observable (e.g., raw text, movie ratings), reported by subjects (e.g., survey responses), or indirectly inferred/derived from other data
(e.g., part-of-speech tags, model-based guesses for age or language)? If the data was reported by subjects or indirectly inferred/derived from other
data, was the data validated/verified? If so, please describe how.

The data was mostly observable from user feedback on feeds recommendation platform of Tencent.

\textbf{What mechanisms or procedures were used to collect the data (e.g.,
hardware apparatuses or sensors, manual human curation, software programs, software APIs)?}How were these mechanisms or procedures validated?

Unknown to the authors of the datasheet.

\textbf{If the dataset is a sample from a larger set, what was the sampling strategy (e.g., deterministic, probabilistic with specific sampling probabilities)?}

we randomly draw users from the database, with the requirement that each user had at least 5 video clicking behaviors.

\textbf{Who was involved in the data collection process (e.g., students,
crowdworkers, contractors) and how were they compensated (e.g., how much were crowdworkers paid)?}

Unknown to the authors of the datasheet.

\textbf{Over what timeframe was the data collected? }Does this timeframe match the creation timeframe of the data associated with the instances (e.g., recent crawl of old news articles)? If not, please describe the timeframe in which the data associated with the instances was created.

We collect user behavior logs from QK/QB from September 17 to December 07, 2021.

\textbf{Did you collect the data from the individuals in question directly, or
obtain it via third parties or other sources (e.g., websites)?}

The data was collected from feeds recommendation platform of Tencent.

\subsection{Preprocessing/cleaning/labeling}

\textbf{Was any preprocessing/cleaning/labeling of the data done (e.g., discretization or bucketing, tokenization, part-of-speech tagging, SIFT feature extraction, removal of instances, processing of missing values)? }If so, please provide a description. If not, you may skip the remaining questions in this section.

We anonymize user ID and item ID to protect user privacy. User profiles are also processed into discrete or binary values.

% At the same time, we discretize the user's age and disrupt the mapping order, which is also to protect user privacy. And process user behavior(i.e. share, follow, like, etc.) into boolean data

\textbf{Was the “raw” data saved in addition to the preprocessed/cleaned/labeled data (e.g., to support unanticipated
future uses)? }If so, please provide a link or other access point to the “raw” data.

No

\subsection{Uses}

\textbf{Has the dataset been used for any tasks already? }If so, please provide
a description.

No

\textbf{Is there a repository that links to any or all papers or systems that
use the dataset? }If so, please provide a link or other access point.

Yes. 

\textbf{What (other) tasks could the dataset be used for?}

The dataset can be used for CTR Prediction, Session-based Recommendation, Mutli-task Learning for Recommendation, Transfer Learning for Recommendation, User Profile Prediction, Lifelong User Representation Learning, Cold-start Recommendation, Model Compression, Model Training Speedup and Model inference Speedup. See our paper for details.

\textbf{Is there anything about the composition of the dataset or the way it
was collected and preprocessed/cleaned/labeled that might impact future uses? }For example, is there anything that a dataset consumer might need to know to avoid uses that could result in unfair treatment of
individuals or groups (e.g., stereotyping, quality of service issues) or other risks or harms (e.g., legal risks, financial harms)? If so, please provide a description. Is there anything a dataset consumer could do to mitigate these risks or harms?

There is little risk here when we have anonymized the dataset.

\subsection{Distribution}

\textbf{Will the dataset be distributed to third parties outside of the entity (e.g., company, institution, organization) on behalf of which the
dataset was created? }If so, please provide a description.

The dataset is publicly available on the internet.

\textbf{How will the dataset will be distributed (e.g., tarball on website, API,
GitHub)? }Does the dataset have a digital object identifier (DOI)?

The distribution of the dataset is detailed in our paper.

\textbf{When will the dataset be distributed?}

The dataset will be distributed in June 2022.

\textbf{Will the dataset be distributed under a copyright or other intellectual
property (IP) license, and/or under applicable terms of use (ToU)? }If so, please describe this license and/or ToU, and provide a link or other access point to, or otherwise reproduce, any relevant licensing terms or
ToU, as well as any fees associated with these restrictions.

This dataset is licensed under a CC BY-NC 4.0 International License(\url{https://creativecommons.org/licenses/by-nc/4.0/}). There is a request to cite the corresponding paper if the dataset is used.

\textbf{Have any third parties imposed IP-based or other restrictions on the
data associated with the instances? } If so, please describe these restrictions, and provide a link or other access point to, or otherwise reproduce,
any relevant licensing terms, as well as any fees associated with these
restrictions.

No

\textbf{Do any export controls or other regulatory restrictions apply to the dataset or to individual instances? }If so, please describe these restrictions, and provide a link or other access point to, or otherwise reproduce, any supporting documentation.

Unknown to authors of the datasset

\subsection{Maintenance}

\textbf{Who will be supporting/hosting/maintaining the dataset?}

Guangnhu Yuan and Fajie Yuan are supporting/maintaining the dataset.

\textbf{How can the owner/curator/manager of the dataset be contacted
(e.g., email address)?}

Guanghu Yuan and Fajie Yuan can be contacted at gh.yuan0@gmail.com and yuanfajie@westlake.edu.cn, respectively.

\textbf{Is there an erratum? }If so, please provide a link or other access point.

Not yet found.

\textbf{Will the dataset be updated (e.g., to correct labeling errors, add new
instances, delete instances)? }If so, please describe how often, by whom, and how updates will be communicated to dataset consumers (e.g., mailing list, GitHub)?

This will be posted on the datasset webpage.

\textbf{Will older versions of the dataset continue to be supported/hosted/maintained? }If so, please describe how. If not, please describe how its obsolescence will be communicated to dataset consumers.

We do not maintain old versions of the dataset, if we update the version of the dataset, we will put the specific details of the dataset update on the relevant GitHub.

\textbf{If others want to extend/augment/build on/contribute to the dataset,
is there a mechanism for them to do so? }If so, please provide a description. Will these contributions be validated/verified? If so, please describe
how. If not, why not? Is there a process for communicating/distributing these contributions to dataset consumers? If so, please provide a description.

If others want to extend/augment/build on/contribute to the dataset, please contact the original authors about incorporating fixes/extensions.

% {
% \small
% \bibliographystyle{unsrt}
% \bibliography{sample.bib}
% }
% {
% \small
% \bibliographystyle{plain}
% \bibliography{rs_benchmark.bib}
% }

\end{document}